\documentclass[a4paper,11pt]{article}

\usepackage{jheppub}
\usepackage{graphicx}
\usepackage{adjustbox}
\usepackage{multirow}
\usepackage{amsmath,amssymb,amsfonts,slashed}
\usepackage{amsthm}
\usepackage{mathrsfs}
\usepackage[title]{appendix}
\usepackage{xcolor}
\usepackage{textcomp}
\usepackage{manyfoot}
\usepackage{booktabs}
\usepackage{algorithm}
\usepackage{algorithmicx}
\usepackage{algpseudocode}
\usepackage{listings}
\usepackage{enumitem}
\usepackage{tikz}
\usepackage[compat=1.1.0]{tikz-feynman}
\usetikzlibrary{shapes, arrows, calc}
\usetikzlibrary{decorations.pathreplacing, calligraphy}
\usepackage{colortbl}
\usepackage{bm}
\usepackage{cprotect}

\newcommand{\lam}{\lambda}
\newcommand{\eps}{\epsilon}
\newcommand{\ol}[1]{\overline{#1}}

\renewcommand{\L}{{\cal L}}
\newcommand{\SU}{{\rm SU}}
\newcommand{\GeV}{{\rm GeV}}

\newcommand{\MeV}{{\rm MeV}}

\newcommand{\eV}{{\rm eV}}

\newcommand{\BNV}{{\rm BNV}}
\newcommand{\yuk}{{\rm yuk}}
\newcommand{\hc}{{\rm h.c.}}

\newcommand{\beq}{\begin{equation}}
\newcommand{\eeq}{\end{equation}}
\newcommand{\bea}{\begin{eqnarray}}
\newcommand{\eea}{\end{eqnarray}}

\renewcommand{\(}{\left(}
\renewcommand{\)}{\right)}
\definecolor{nicered}{rgb}{0.7,0.1,0.1}
\definecolor{nicegreen}{rgb}{0.1,0.5,0.1}
\definecolor{red}{rgb}{1.0, 0, 0}
\definecolor{niceblue}{rgb}{0,0,0.8}
\definecolor{red}{rgb}{1.0, 0, 0}
\hypersetup{colorlinks,bookmarksopen,bookmarksnumbered,
linkcolor=blus,pdfstartview=FitH,urlcolor=rossos,citecolor=verde}
\allowdisplaybreaks
\hypersetup{colorlinks,citecolor= nicegreen,linkcolor= niceblue,urlcolor=nicered}

\graphicspath{{figs/}} 

\begin{document}

\preprint{
\begin{flushright}
YITP-26-10\\
KYUSHU-HET-350
\end{flushright}
}

\title{UV cut-off of the Standard Model and proton decays}

\author[a]{Ryuichiro Kitano,}
\author[b,c,d]{Shohei Okawa}

\affiliation[a]{Yukawa Institute for Theoretical Physics, Kyoto University, Kyoto 606-8502 Japan}
\affiliation[b]{Asia Pacific Center for Theoretical Physics, Pohang, 37673, Korea}
\affiliation[c]{Department of Physics, Pohang University of Science and Technology, Pohang, 37673, Korea}
\affiliation[d]{Department of Physics, Kyushu University, 744 Motooka, Nishi-ku, Fukuoka, 819-0395, Japan}

\emailAdd{ryuichiro.kitano@yukawa.kyoto-u.ac.jp}
\emailAdd{shohei.okawa@apctp.org}

\abstract{
Non-observation of proton decays as well as the smallness of the neutrino masses can naturally be explained by the accidental baryon and lepton number symmetry in the Standard Model, where the approximate symmetries are a consequence of the absence of the baryon or lepton number violating operators at the renormalizable level.
The neutrino masses at sub-eV scales can be explained by the presence of the dimension-five term, $\ell\ell HH/\Lambda$, in the Lagrangian, suggesting that a more fundamental theory takes over beyond the energy scale $\Lambda$.
We consider the possibility that the theory above the scale $\Lambda$ generates general higher dimensional operators with the flavor structure implied by the Yukawa interactions in the Standard Model. Such a set-up can be realized, for example, in the composite Higgs scenario with partial compositeness of fermions. The fermion masses and the neutrino masses are explained for $\Lambda \gtrsim 10^{11}$~GeV.
The lifetime of proton in this scenario is, interestingly, consistent with the observed event of the $p \to \pi^0 \mu^+$ decay at the Super-Kamiokande experiment.
The Hyper-Kamiokande experiment should see a large number of events soon after the data taking, if the event observed by the Super-Kamiokande is due to the real proton decay.
}

\maketitle


\section{Introduction}

Various observations, such as neutrino masses, dark matter, the baryon asymmetry of the Universe, and inflation, as well as the necessity of quantum gravity, suggest the existence of physics beyond the Standard Model.
The scale of new physics is, however, not predicted within the Standard Model because of its renormalizability, unless naturalness arguments are invoked.
In the minimalist approach, the next scale beyond the electroweak scale is the one with 
the dimension-five operators, $\ell\ell HH/\Lambda$, to explain the 
observed neutrino oscillation phenomena~\cite{Weinberg:1979sa}.
The Standard Model added by the $\ell\ell HH$ term has an UV cut-off scale, $\Lambda$, at which we expect some new physics to 
replace the description.
Of course, the simplest theory beyond $\Lambda$ is the appearance
of the right-handed neutrinos~\cite{Minkowski:1977sc,Yanagida:1979as,Gell-Mann:1979vob,Mohapatra:1979ia}.
One can, on the other hand, consider an extreme possibility that
$\Lambda$ is really the UV cut-off scale; the particle descriptions of (some of) the Standard Model particles are lost beyond $\Lambda$ and there is no lepton number conservation in the full theory so that the effective 
description contains the $\ell\ell HH$ operator. In that case,
we expect that many kinds of higher dimensional operators appear
with the suppression of $1/\Lambda^n$ with $n$ the appropriate powers 
which depend on the dimension of the operators. The neutrino masses
are the largest consequence of this theory as the dimension-five
operator, $\ell\ell HH$, is the leading non-renormalizable operator.

The structure of the Yukawa interactions of fermions in the Standard Model
suggests that the fermions are not quite strongly coupled to the
dynamics which appears at $\Lambda$, as the Yukawa interactions are 
quite weak except for the top quark. Even for the top quark,
it is ${\cal O}(1)$ at the electroweak scale and gets weaker at high energy. 
Interestingly, the paradigm of the ``partial compositeness''~\cite{Kaplan:1991dc} provides 
us with a consistent framework to explain the flavor structures.
There is a fact that the Cabibbo-Kobayashi-Maskawa (CKM) and Pontecorvo-Maki-Nakagawa-Sakata (PMNS) matrices have a common feature
that the generations with large mass hierarchies have small mixings.
This can be naturally realized by the $\epsilon$ scheme that
the Yukawa couplings as well as the neutrino masses have a structure of $Y_{u,d}^{(ij)} \sim \epsilon_q^{(i)} \epsilon_{u,d}^{(j)}$, $Y_\ell \sim \epsilon_\ell^{(i)} \epsilon_e^{(j)}$, and $m_\nu^{(ij)} \propto \epsilon_\ell^{(i)} \epsilon_\ell^{(j)}$,
where $\epsilon$'s are suppression factors to represent the ``compositeness'' of each fermion operator, $q, u, d, \ell$, and $e$.
In order to obtain the ${\cal O}(1)$ Yukawa coupling of the top quark, the Higgs
field is necessarily composite, at least largely.
One popular example is to identify the Higgs field as the pseudo Nambu-Goldstone boson of the strong dynamics~\cite{Kaplan:1983fs,Kaplan:1983sm,Georgi:1984ef,Georgi:1984af,Dugan:1984hq,Contino:2003ve}. The small quartic coupling
constant of the Higgs potential can also be explained
by $\epsilon$ factors in that case.

Combining the known facts, the finite neutrino masses and the flavor structure,
one can consider a minimalist scenario. There appears a strong dynamics
(or possibly the theory of quantum gravity) at the scale $\Lambda$ in which all the global symmetries in the Standard Model
are maximally broken, and through the mixing between the elementary
field operators and fully composite ones, 
one obtains the dimension-four Yukawa interactions as well as the dimension-five $\ell\ell HH$ operators to explain all the known facts.

This simple scenario needs to pass some consistency checks.
We would expect to have further higher dimensional operators such as  dimension-six operators to induce proton decays. The lifetime needs 
to be long enough to be consistent with the experimental limits. This should give a lower bound on the scale $\Lambda$.
Also, the Higgs boson should appear as a light degree of freedom.
We do not try to solve the hierarchy problem 
between $\Lambda$ and the electroweak scale, but 
at least the coefficient of the quartic potential term should be small.
This means that the Higgs field parametrizes an approximately flat direction
of the theory such as pseudo Nambu-Goldstone bosons
or flat directions in the supersymmetric theories.
In many cases, the quartic coupling is predicted to be positive
at the scale $\Lambda$, as otherwise the flat direction gets unstable.
Since the running of the Standard Model parameter
shows that the quartic term drives into negative above some intermediate
scale, this discussion provides an upper bound on $\Lambda$ as $\Lambda\lesssim{\cal O}(10^{11})\,\GeV$~\cite{Degrassi:2012ry}.

In this paper,
we perform a consistency test on this scenario by evaluating proton decays induced from dimension-six baryon-number-violating operators. 
The scale and the flavor structure of the relevant baryon-number-violating interactions are determined by the seesaw formula and the $\epsilon$ parameterization suggested from the Yukawa couplings and neutrino masses. 
The partial decay rates are therefore characterized by the $\epsilon$ factors.
It naturally suppresses proton decays into the first-generation leptons and predicts the leading decay modes associated with the second-generation leptons, such as $p \to \pi^0 \mu^+$ and $p\to K^0 \mu^+$.
Varying a few model parameters, we show the proton lifetimes are minimized at $\Lambda\sim10^{11}$\,GeV, a lowest scale to reproduce the tau mass, and for $\Lambda\sim10^{11}$\,GeV, 
the allowed region to satisfy the limit of the proton lifetime from the Super-Kamiokande experiment
starts to open~\cite{Super-Kamiokande:2005lev, Super-Kamiokande:2012zik, Super-Kamiokande:2013rwg, Super-Kamiokande:2014otb, Super-Kamiokande:2017gev, Super-Kamiokande:2020wjk, Super-Kamiokande:2022egr, Super-Kamiokande:2024qbv, Super-Kamiokande:2025ibz}.
A large parameter region can be tested by the future Hyper-Kamiokande experiment~\cite{Hyper-Kamiokande:2018ofw}.

This paper is organized as follows. 
In Sec.~\ref{sec:eps_scheme}, we overview the Yukawa structure and $\epsilon$ parameterization suggested from the observed fermion masses and mixings.
In Sec.~\ref{sec:lifetime}, we outline the methodology of proton decay calculation and provide analytical approximation of the proton lifetimes. 
The minimum proton lifetimes within the $\epsilon$ ansatz are also shown there, together with the current experimental limits.
We summarize our findings in Sec.~\ref{sec:summary}. 
Analytical expressions and technical details of the calculation are summarized in the appendices.

\section{Yukawa structure and $\epsilon$ scheme}
\label{sec:eps_scheme}

The CKM and PMNS matrices as well as
the mass hierarchy of the fermions
have a suggestive feature in the Standard Model.
As is well-known in flavor model-building, the structure can be naturally explained by the ansatz
that each fermion field in the Standard Model, $q$, $u$, $d$, $\ell$ and $e$,
has its own suppression factors, $\epsilon_q$, $\epsilon_u$, $\epsilon_d$, $\epsilon_\ell$,
and $\epsilon_e$ for each generation.
The suppression factors appear in the Yukawa interactions so that 
the mass hierarchies are explained. This scheme predicts that
the off-diagonal components of the Yukawa matrices are related
to the hierarchy in the diagonal entries, and thus gives a non-trivial
structure in the quark and lepton mixing matrices.
Amazingly, the Standard Model indeed has such a structure. The epsilon factors
are a natural consequence of various scenarios, such as the Froggatt-Nielsen model~\cite{Froggatt:1978nt}, wave function profiles in the extra dimension as well as partial compositeness models~\cite{Huber:2003tu,Csaki:2008zd}.
In the case of partial compositeness, the $\epsilon$ factors represent the 
degree of ``compositeness,'' where the degrees are controlled by couplings between
the elementary fields and the operators in the strongly interacting sector.
The maximum value $\epsilon = 1$ means that the fermion is a member of the
strong sector, i.e., a composite particle, while $\epsilon < 1$ means that
the coupling is suppressed by $\epsilon$ compared to the natural size in the 
strong dynamics.
As is well-known, the very same parametrization works for the overlap of the 
wave function profiles among fermions and the Higgs field in the extra-dimensions.
We take the picture of the partial compositeness and determine the $\epsilon$ factors
from the mass hierarchies and mixings of quarks and leptons, and apply the same rule to
the operators that induce proton decays.

The grand picture is thus the presence of the UV cut-off scale, $\Lambda$, 
at which some
strong dynamics (such as quantum gravity) appears, and all sorts of operators consistent with the gauge invariance in the low energy effective theory are 
generated by the dynamics with the rule that the interactions 
of each fermion are suppressed by the $\epsilon$ factors.
We do not try to explain the weakness of gravity, $\Lambda \ll M_{\rm Pl}$, and the light Higgs boson, $m_h \ll \Lambda$,
in the following discussion. 

\subsection{Ansatz on the Yukawa structures}

The quark and lepton masses and mixings are generated from the following interaction Lagrangian, 
\beq
{\cal L}_{\rm mass} = - f_d^{ij} H^\dagger q_{i} \bar{d}_{j} - f_u^{ij} H q_{i} \bar{u}_{j} - f_e^{ij} H^\dagger \ell_{i} \bar{e}_{j} - \frac{\lam_{ij}}{\Lambda} (\ell_i H)(\ell_j H) + \hc 
\eeq
Here and throughout this paper, we employ the two-component Weyl spinor notation, following the convention of ref.~\cite{Dreiner:2008tw} (see also refs.~\cite{Beneito:2023xbk,Gargalionis:2024nij}). 
Then, $q, \ell$ denote the left-handed quarks and leptons, and 
$\bar u^\dagger, \bar d^\dagger, \bar e^\dagger$ the right-handed quarks and leptons. 
The quark and lepton mass matrices read 
\begin{align}
(M_u)_{ij} & = \frac{f_u^{ij}v}{\sqrt{2}} \,,\quad
(M_d)_{ij} = \frac{f_d^{ij}v}{\sqrt{2}} \,,\\
(M_e)_{ij} & = \frac{f_e^{ij}v}{\sqrt{2}}  \,,\quad
(M_\nu)_{ij} = \frac{\lam^{ij}v^2}{\Lambda}  \,.
\end{align}
with $v=246\,\GeV$.
Here, we assume the following flavor structure for the interaction matrices, 
\beq
f_{u}^{ij} \sim \lam_\yuk^{ij} \, \eps_{q_i} \eps_{u_j} \,, \quad 
f_{d}^{ij} \sim \lam_\yuk^{ij} \, \eps_{q_i} \eps_{d_j} \,, \quad 
f_{e}^{ij} \sim \lam_\yuk^{ij} \, \eps_{\ell_i} \eps_{e_j} \,, \quad 
\lam_{ij} \sim (\lam_\yuk^2)^{ij} \, \eps_{\ell_i} \eps_{\ell_j} \,,
\eeq
where 
$|\eps_f| \leq 1$ ($f=q,u,d,\ell,e$) and 
$\lambda_\yuk$'s are $3\times3$ democratic matrices whose entries are of ${\cal O}(4 \pi)$ or less and have no hierarchies.
These matrices are generically all different for each Yukawa matrix ($f_u$, $f_d$, $f_e$ and $\lambda$) and represent
random matrices. Below, we sometimes denote $\lambda_{\rm yuk}$ as the typical size of the entries.
In the case of a large $N$ theory
for composite Higgs, it is typically of ${\cal O}(4 \pi / \sqrt N)$, where $N$ denotes the number of ``colors''.
With this ansatz, the structure of $\eps_{q_i}$ parameters are determined from the quark mixing, \beq
\eps_{q_i} \propto V_{i3} \simeq (0.004, 0.04,1) \,,
\label{eq:eps1}
\eeq
where $V_{ij}$ denotes the CKM matrix elements.\footnote{Here, the symbol `$\propto$' in the equation does not means the exact proportionality, but that $V_{i3}$ represents the typical flavor hierarchy of $\epsilon_{q_i}$ up to ${\cal O}(1)$ uncertainty.
In general one can, for example, express it as $\epsilon_{q_i} = \epsilon_{q_3} \frac{c_i}{c_3} \frac{|V_{i3}|}{|V_{33}|}$ with $c_i$ being random numbers of ${\cal O}(1)$. 
The parameterization we will use in this paper is given in Sec.~\ref{sec:parametrization}.} 
Given no hierarchical structure in $\lambda_{\rm yuk}$'s, the above setting of $\epsilon_{q_i}$
induces the correct mixing angles in the CKM matrix
for the third-generation quarks.
It is already non-trivial that $\eps_{q_1} / \eps_{q_2} \sim 0.1$ is consistent with 
the Cabibbo angle, $\lambda \sim 0.2$.
The quark masses then determine
the $\eps_{u_i}$ and $\eps_{d_i}$ factors:
\begin{align}
\eps_{u_i} & \propto m_{u_i}/V_{i3} \propto (0.004, 0.2, 1) \,,\\
\eps_{d_i} & \propto m_{d_i}/V_{i3} \propto (0.2, 0.6, 1)  \,.
\end{align}
Similarly, the lepton mixing matrix $U$ and 
the charged lepton masses can determine 
the structures of $\eps_{\ell_i}$ and $\eps_{e_i}$
such as
\begin{align}
\eps_{\ell_i} & \propto U_{i3} \simeq (0.15, 0.7, 0.7) \,,\\
\eps_{e_i} & \propto m_{e_i}/\eps_{\ell_i} \propto (0.001, 0.06, 1) \,.
\label{eq:eps2}
\end{align}
This gives the prediction of the 
neutrino mass hierarchy, 
\beq
m_{\nu_i} \propto (\eps_{\ell_i})^2 \propto (0.04, 1, 1) \,.
\eeq
Again, it is consistent with the hierarchy
in the solar and atmospheric neutrino oscillation phenomena, $\Delta m^2_{\rm sol} / \Delta m^2_{\rm atm} \sim 0.03$,
as well as the solar mixing angle, $\sin \theta_{\rm sol} \sim 0.6$,
for the normal ordering of the neutrino
masses
by taking a slightly smaller value of $\eps_{\ell_2}$ or assuming mild cancellation 
in the eigenvalue $m_{\nu_2}$.

Up to overall factors, one could now determine
the hierarchy structures of $\eps$ by the 
data of the fermion masses and mixings.
We emphasize here that the obtained structure 
is independent of microscopic assumptions
except for the Majorana nature of 
neutrino masses.

\subsection{Parametrization}
\label{sec:parametrization}

By fixing the relative sizes of $\eps$, 
one can parametrize the absolute values
of $\eps$ by using $\Lambda$, $\eps_{q_3}$ and $\lam_{\yuk}$.
Here we explicitly show the parameterization of $\eps$, 
which will be used when calculating proton lifetimes in Sec.~\ref{sec:lifetime}.

First, 
$\eps_{q_i}$ is parameterized using the CKM matrix elements:
\beq
\eps_{q_i} = \eps_{q_3} \(\frac{|V_{13}|}{|V_{33}|}, \frac{|V_{23}|}{|V_{33}|}, 1\) \,.
\label{eq:eps_qi}
\eeq
Then, $\eps_{u_i}$ and $\eps_{d_i}$ are determined through the up-type and down-type quark masses respectively:
\begin{align}
\eps_{q_i} \eps_{u_i} \lam_\yuk & = \frac{\sqrt{2}m_{u_i}}{v} 
	~~\Rightarrow~~ \eps_{u_i} \simeq \eps_{q_3}^{-1} \lam_\yuk^{-1} \(\frac{m_{u_i}}{m_t} \frac{|V_{33}|}{|V_{i3}|}\) \,,\label{eq:eps_ui}\\
\eps_{q_i} \eps_{d_i} \lam_\yuk & = \frac{\sqrt{2}m_{d_i}}{v} 
	~~\Rightarrow~~ \eps_{d_i} \simeq 0.024 \, \eps_{q_3}^{-1} \lam_\yuk^{-1} \(\frac{m_{d_i}}{m_b} \frac{|V_{33}|}{|V_{i3}|}\) \,.\label{eq:eps_di}
\end{align}
All $\eps$ in the quark sector are thus expressed only by $\eps_{q_3}$ and $\lambda_\yuk$.
Imposing $|\eps_{u_3}|\leq1$ and fitting the top quark mass, 
the values of $\eps_{q_3}$ and $\lambda_\yuk$ are restricted in the range, 
\beq
|\eps_{q_3}|\leq1\,,\quad
\lambda_\yuk\leq4\pi\,,\quad
1\leq|\eps_{q_3}\lambda_\yuk|\leq4\pi\,.
\label{eq:eps-lam_range}
\eeq
As for leptons, $\eps_{\ell_i}$ is parameterized using the PMNS matrix elements:
\beq
\eps_{\ell_i} = \eps_{\ell_3} \( \frac{|U_{13}|}{|U_{33}|}, \frac{|U_{23}|}{|U_{33}|}, 1\) \,.
\eeq 
$\eps_{\ell_3}$ is fixed by the heaviest neutrino mass $m_3 \simeq \sqrt{\Delta m_{31}^2} = 0.05\,\eV$, 
\beq
m_3 = \frac{(\lam_\yuk \eps_{\ell_3})^2 v^2}{\Lambda} 
	~~\Rightarrow~~ \eps_{\ell_3} = \frac{\sqrt{m_3\Lambda/v^2}}{\lam_\yuk} 
	\simeq 0.029 \, \lam_\yuk^{-1} \, \sqrt{\frac{\Lambda}{10^{12}\,\GeV}} \,.
\label{eq:epsL-seesaw}
\eeq
Furthermore, $\eps_{e_i}$ is determined by the lepton masses and eq.~(\ref{eq:epsL-seesaw}):
\beq
\eps_{\ell_i} \eps_{e_i} \lam_\yuk = \frac{\sqrt{2}m_{\ell_i}}{v} 
	~~\Rightarrow~~ \eps_{e_i} \simeq 0.36 \, \(\frac{m_{\ell_i}}{m_\tau} \frac{|U_{33}|}{|U_{i3}|}\) \sqrt{\frac{10^{12}\,\GeV}{\Lambda}} \,.
\label{eq:eps_ei}
\eeq
where we note $\eps_{e_i}$ is independent of $\lam_\yuk$ and determined solely by $\Lambda$.
Imposing $|\eps_{e_3}| \leq 1$ in eq.~(\ref{eq:eps_ei}), 
a lower limit on the scale $\Lambda$ is obtained: 
\beq
\Lambda \gtrsim 1.3 \times 10^{11}\,\GeV\,.
\label{eq:Lambda_min}
\eeq

\subsection{GUT relation}

From eqs.~(\ref{eq:eps1})-(\ref{eq:eps2}), 
one can see that $\eps$'s can be grouped into two hierarchical patterns: 
a large hierarchy ($\epsilon_q$, $\epsilon_u$, $\epsilon_e$) and a moderate hierarchy ($\epsilon_d$, $\epsilon_\ell$).
This grouping is reminiscent of matter unification in the $\SU(5)$ grand unified theory (GUT).
Here, inspired by this fact, let us mention the possibility that $\epsilon$'s are unified like $\epsilon_q\sim\epsilon_u\sim\epsilon_e$ and $\epsilon_{d}\sim\epsilon_\ell$.

First, by imposing $\epsilon_{q_3}\sim\epsilon_{u_3}$ in eq.~(\ref{eq:eps_ui}), we obtain 
\beq
\epsilon_{q_3} \sim \epsilon_{u_3} \sim 0.99\,\lambda_{\yuk}^{-1/2} \,.
\label{eq:q-u}
\eeq
Then, by equating $\epsilon_{q_3}$ to eq.~(\ref{eq:eps_ei}) with $i=3$, we find
\beq
\epsilon_{q_3} \sim \epsilon_{e_3} \sim 1.1 \sqrt{\frac{10^{11}\,{\rm GeV}}{\Lambda}} \,.
\label{eq:GUT1}
\eeq
Thus, the unification $\epsilon_q\sim\epsilon_u\sim\epsilon_e\sim0.99\,\lambda_{\yuk}^{-1/2}$ is achieved at the scale
\beq
\Lambda \sim 1.3\times 10^{11}\,{\rm GeV} \times \lambda_{\rm yuk} \,.
\eeq
Similarly, the unification $\epsilon_{d}\sim\epsilon_{\ell}$ is realized by equating eq.~(\ref{eq:epsL-seesaw}) to eq.~(\ref{eq:eps_di}) with $i=3$, which yields
\beq
\epsilon_{q_3} \sim 2.6 \sqrt{\frac{10^{11}\,{\rm GeV}}{\Lambda}} \,.
\label{eq:GUT2}
\eeq
Interestingly, the two unification relations, eqs.~(\ref{eq:GUT1}) and (\ref{eq:GUT2}), are satisfied approximately at the same scale.
Note that in the above calculation, 
we do not use the Yukawa couplings evaluated at high-energy scales $\Lambda\sim10^{11}\,{\rm GeV}$.
Once the running of the Yukawa couplings is taken into account, 
agreement between eqs.~(\ref{eq:GUT1}) and (\ref{eq:GUT2}) is improved significantly.

With the two GUT-inspired relations, we only have one free parameter. 
For example, when $\epsilon_{q_3}$ is chosen to be a free parameter, 
the other two parameters, $\lambda_{\rm yuk}$ and $\Lambda$, are determined by 
$\lambda_{\rm yuk} \sim (\epsilon_{q_3})^{-2}$ and 
$\Lambda \sim 1.3 \times 10^{11}\,{\rm GeV} \times (\epsilon_{q_3})^{-2}$.

\section{Proton lifetimes}
\label{sec:lifetime}

With the parametrization discussed in the previous section, 
we study the proton lifetimes by considering general dimension-six operators that violate the baryon number.

\subsection{SMEFT operators with $\Delta B=\Delta L=1$}

Below the scale $\Lambda$, 
the relevant dynamical degrees of freedom are the Standard Model fields, and physics is described by the Standard Model Effective Field Theory (SMEFT) \cite{Buchmuller:1985jz, Grzadkowski:2010es}. 
In this case, the leading contribution to proton decays is expected to arise from dimension-six baryon-number-violating (BNV) operators with $\Delta B=\Delta L=1$, 
\beq
\L_{\rm BNV}^{d=6} = \sum_I \sum_{i,jk,l} C_{I,\,ijkl}\,{\cal O}_{I,\,ijkl} \,,
\eeq
where a complete set of the BNV operators ${\cal O}_{I,\,ijkl}$ is given by \cite{Grzadkowski:2010es}
\begin{align}
{\cal O}_{qqq\ell,\,ijkl} & =  (q^\alpha_i q^\beta_j) (q^\gamma_k \ell^\delta_l) \eps_{\alpha\gamma} \eps_{\beta\delta} \\
{\cal O}_{qque,\,ijkl} & = (q^\alpha_i q^\beta_j) (\bar{u}_k^\dag \bar{e}_l^\dag) \eps_{\alpha\beta} \\
{\cal O}_{duue,\,ijkl} & = (\bar{d}_i^\dag \bar{u}_j^\dag) (\bar{u}_k^\dag \bar{e}_l^\dag) \\
{\cal O}_{duq\ell,\,ijkl} & =  (\bar{d}_i^\dag \bar{u}_j^\dag) (q^\alpha_k \ell^\beta_l) \eps_{\alpha\beta} 
\end{align}
where $\alpha,\beta,\gamma,\delta=1,2$ denote the $\SU(2)_L$ indices and $i,j,k,l=1,2,3$ the quark and lepton flavor indices. 
The QCD indices are suppressed here. 

We apply the $\eps$ parameterization to describe the flavor structure of the BNV interactions. 
Then, the Wilson coefficients for the proton decay operators are expressed as
\begin{align}
C_{qqql,\,ijkl} & =  \lam_\BNV^2 \frac{c_{qqq\ell}}{\Lambda^2} \eps_{q_i} \eps_{q_j} \eps_{q_k} \eps_{\ell_l} \,, \qquad
C_{qque,\,ijkl} = \lam_\BNV^2 \frac{c_{qque}}{\Lambda^2} \eps_{q_i} \eps_{q_j} \eps_{u_k} \eps_{e_l} \,,\\
C_{duue,\,ijkl} & = \lam_\BNV^2 \frac{c_{duue}}{\Lambda^2} \eps_{d_i} \eps_{u_j} \eps_{u_k} \eps_{e_l} \,, \qquad~
C_{duql,\,ijkl} = \lam_\BNV^2 \frac{c_{duq\ell}}{\Lambda^2} \eps_{d_i} \eps_{u_j} \eps_{q_k} \eps_{\ell_l} \,, 
\end{align}
where $\lam_\BNV$ is a UV-model dependent parameter and $\lam_\BNV=\lam_\yuk=4\pi/\sqrt{N}$ is expected in composite Higgs scenarios.\footnote{There are several works analyzing proton decays in supersymmetric models and SMEFT with the Minimal Flavor Violation (MFV)~\cite{Chivukula:1987py, DAmbrosio:2002vsn} or ${\rm U}(2)^5$ \cite{Barbieri:2011ci, Barbieri:2012uh, Blankenburg:2012nx} structures rather than the $\epsilon$ parameterization. 
See refs.~\cite{Nikolidakis:2007fc, Smith:2011rp, Helset:2019eyc, Beneito:2025fzf} for example. 
The flavor structure in the $\epsilon$ scheme is in some sense understood as square-root of the MFV structure where the flavor mixing is governed by the $\epsilon$ factors rather than full Yukawa matrices.} 
Here, $c_{qqq\ell}$, $c_{qque}$, $c_{duue}$, and $c_{duq\ell}$ in general represent ${\cal O}(1)$ parameters with the flavor indices.

\subsection{Methodology for proton decay calculation}

Nucleon decays through the dimension-six BNV SMEFT interactions are well-studied and can be systematically evaluated \cite{Nath:2006ut, Jenkins:2017jig, Jenkins:2017dyc, Beneito:2023xbk, Gargalionis:2024nij}. 
First, the Wilson coefficients of the BNV operators are evolved from $\mu=\Lambda$ to $\mu = m_W$ by solving the renormalization group (RG) equations within the SMEFT \cite{Alonso:2014zka, Banik:2025wpi}.  
The RG-improved SMEFT Lagrangian is matched at $\mu = m_W$ to the Low Energy Effective Field Theory (LEFT) \cite{Jenkins:2017jig}, where the electroweak symmetry is broken and the dynamical degrees of freedom consist of photon, gluon, the five quarks and all of the neutral and charged leptons. 
The threshold corrections at the matching scale are also taken into account accordingly \cite{Dekens:2019ept}. 

Below the electroweak scale, physics is governed by the LEFT Lagrangian. 
The RG effects to the hadronic scale $\mu=\Lambda_{\rm QCD}$ (typically $\Lambda_{\rm QCD}=2\,\GeV$) are calculated using anomalous dimensions of the LEFT operators, together with the RG equations for the quark and lepton masses and the QCD and QED gauge couplings, which are modified by the inclusion of the pertinent LEFT interactions \cite{Abbott:1980zj, Jenkins:2017dyc, Aebischer:2025hsx, Naterop:2025lzc}. 
Although a complete set of the one-loop anomalous dimensions is known within the LEFT, the RG effects are dominated by the QCD interactions. 
During the RG calculation to $\mu = \Lambda_{\rm QCD}$, one needs to include corrections appropriately at each of the bottom and charm quark mass thresholds. 

With the Wilson coefficients of the BNV LEFT operators at $\mu=\Lambda_{\rm QCD}$ in hand, 
the evaluation of nucleon decay rates into a meson and a lepton requires non-perturbative determinations of the nucleon-to-meson transition matrix elements. 
This computation can be carried out by using Baryon Chiral Perturbation Theory (BChPT) \cite{JLQCD:1999dld, Aoki:2008ku} or by lattice QCD simulation \cite{Aoki:2008ku, Yoo:2021gql, Bali:2022qja}, when combined with the lattice-to-$\overline{\rm MS}$ matching \cite{Pivovarov:1991nk, Aoki:2006ib, Gracey:2012gx}. 

In this work, we follow the same methodology for the proton lifetime calculation as outlined in the literature \cite{Beneito:2023xbk, Gargalionis:2024nij}, except that we ignore the RG corrections from the cut-off scale $\Lambda$ to the experimental scale $\mu_{\rm exp}\simeq\,\GeV$, since these corrections do not change the order of the interaction strength, nor modify the flavor structure of the BNV interactions. 
All details for the calculation and the hadronic inputs we use in this paper are summarized in the appendices.

\subsection{Analysis}

There are three free parameters in our framework:  
$\eps_{q_3}$, $\Lambda$, and $\lambda_\yuk$. 
The rest of $\eps$ can be expressed only by these parameters, up to ${\cal O}(1)$ uncertainty. 
In this section, 
we provide analytical approximations for the proton lifetimes. 
Contributions from four BNV operators can interfere each other in general. 
However, we are only interested in the scaling of each contribution with $\eps_{q_3}$, $\Lambda$, and $\lambda_\yuk$, and the interference is ignored in our analysis. 
Below, we consider the large $N$ theory of composite Higgs scenarios where $\lam_\BNV=\lam_\yuk=4\pi/\sqrt{N}$.

As mentioned in Sec.~\ref{sec:parametrization}, fitting the tau mass requires $\Lambda \gtrsim 1.3\times10^{11}$~GeV. 
If we also impose the positivity of the Higgs quartic coupling, 
the cut-off scale $\Lambda$ is bounded from above \cite{Degrassi:2012ry}.
According to the state-of-the-art analysis of the Higgs effective potential in the Standard Model, 
the positivity requires $\Lambda \lesssim 5.3 \times 10^{11}$~GeV \cite{Hiller:2024zjp} when using the PDG2024 central values for the top, Higgs, and $Z$ pole masses and the 5-flavor strong coupling $\alpha_s^{(5)}(\mu=M_Z)$.
Therefore, $10^{11}\,\GeV \lesssim \Lambda \lesssim 5\times10^{11}\,\GeV$ is a well-motivated window of the cut-off scale in this scenario.
Below, we will study whether the proton lifetime predicted within this scale window is consistent with the current experimental results and is within the reach of the future experiments.

\begin{table}[tb]
\centering
\begin{tabular}{|c|c|c|c|c|}
\hline
Decay mode &  $a_{qqq\ell}$ [$10^{34}\,{\rm yrs}$] & $a_{qque}$ [$10^{34}\,{\rm yrs}$] & $a_{duue}$ [$10^{34}\,{\rm yrs}$] & $a_{duq\ell}$ [$10^{34}\,{\rm yrs}$] \\
\hline
~$p \to \pi^0 e^+$~ & ~$2.1$~ & ~$1.0$~ & ~$7.1\times10^{-3}$~ & ~$95$~ \\ \hline
$p \to \pi^+ \ol{\nu}_e$ & $1.1$ & 0 & 0 & $48$ \\ \hline
$p \to K^0 e^+$ & $6.6\times10^{-3}$ & $2.2$ & $1.0\times10^{-3}$ & $9.0\times10^{3}$ \\ \hline
$p \to K^+ \ol{\nu}_e$ & $3.0\times10^{-2}$ & 0 & 0 & $1.5$ \\ \hline
$p \to \eta e^+$ & $14$ & $1.2$ & $9.0\times10^{-2}$ & $1.1\times10^{2}$ \\ \hline
\hline
~$p \to \pi^0 \mu^+$~ & ~$9.8\times10^{-2}$~ & ~$5.3\times10^{-4}$~ & ~$3.6\times10^{-6}$~ & ~$4.4$~ \\ \hline
$p \to \pi^+ \ol{\nu}_\mu$ & $4.9\times10^{-2}$ & 0 & 0 & $2.2$ \\ \hline
$p \to K^0 \mu^+$ & $3.0\times10^{-4}$ & $1.1\times10^{-3}$ & $0.50\times10^{-6}$ & $4.1\times10^{2}$ \\ \hline
$p \to K^+ \ol{\nu}_\mu$ & $1.4\times10^{-3}$ & 0 & 0 & $7.0\times10^{-2}$ \\ \hline
$p \to \eta \mu^+$ & $0.62$ & $6.1\times10^{-4}$ & $4.6\times10^{-5}$ & $5.1$ \\ \hline
\end{tabular}
\caption{
Numerical values of $a_{qqq\ell}$, $a_{qque}$, $a_{duue}$ and $a_{duq\ell}$ for proton decays.
The values correspond to the minimum proton lifetimes induced from each BNV operator. 
See eq.~(\ref{eq:Gamma_scaling}) for the definition.
}
\label{tab:Gamma_P}
\end{table}

Applying our parameterization, the inverse of the partial proton decay rate is given in the form
\begin{align}
\Gamma^{-1}
	& \sim a_{qqql}\,\frac{\Lambda_{11}^3}{c_{qqq\ell}^2 \, \eps_{q_3}^6} \(\frac{4\pi}{\lam_\yuk}\)^{2}
			+ a_{qque}\,\frac{\Lambda_{11}^5}{c_{qque}^2 \, \eps_{q_3}^2} \(\frac{4\pi}{\lam_\yuk}\)^{2} \nonumber\\
	& \quad + a_{duue}\,\frac{\Lambda_{11}^5}{c_{duue}^2} \(\frac{\eps_{q_3}}{1/4\pi}\)^6 \(\frac{\lam_\yuk}{4\pi}\)^2
			+ a_{duq\ell}\,\frac{\Lambda_{11}^3}{c_{duq\ell}^2} \(\frac{\eps_{q_3}}{1/4\pi}\)^2 \(\frac{\lam_\yuk}{4\pi}\)^2 \,,
\label{eq:Gamma_scaling}
\end{align}
where we define $\Lambda_{11} := \Lambda/(10^{11}\,\GeV)$ and $a_{qqql}, a_{qque}, a_{duue}$, and $a_{duq\ell}$ are numerical coefficients for each decay mode given in Table \ref{tab:Gamma_P}.
The symbol $\sim$ in the equation means that the interference among the different BNV operators is neglected. 
As the neutrinos are left-handed, decay modes involving $\overline\nu_{e,\mu,\tau}$ in the final state are induced only from the $qqq\ell$ or $duq\ell$ operators.
The neutron decay rates are closely related to those of the proton:
\begin{align}
\Gamma^{-1}(n \to \pi^- \ell^+) 
	& \simeq 3\,\Gamma^{-1}(p \to \pi^0 \ell^+) \nonumber\\
\Gamma^{-1}(n \to \pi^0 \bar{\nu}) 
	& \simeq \frac{1}{2}\,\Gamma^{-1}(p \to \pi^+ \ol{\nu}) \label{eq:Gamma_n}
\end{align}

There is a general scaling rule of the inverse decay rate $\Gamma^{-1}$ with $\eps_{q_3}$, $\Lambda_{11}$ and $\lambda_\yuk$. 
Firstly, we find $\Gamma^{-1} \propto \eps_{q_3}^{-2(n-m)}$, 
where $n$ is the number of $q$ fields and $m$ is the total number of $u$ and $d$ fields in the corresponding BNV operators. 
Secondly, the scaling with $\Lambda_{11}$ is determined solely by the type of lepton field involved in the BNV operators. 
The inverse decay rate scales as $\Gamma^{-1} \propto \Lambda_{11}^3$ or $\Lambda_{11}^5$, 
depending on whether the lepton is left-handed or right-handed. 
This is understood from $\eps_{\ell_i} \propto \Lambda^{1/2}$ and $\eps_{e_i} \propto \Lambda^{-1/2}$.
Lastly, the $\lambda_\yuk$-scaling follows $\Gamma^{-1} \propto \lambda_\yuk^{2(x-2)}$, 
where $x$ is the total number of $u$, $d$ and $\ell$ fields in the BNV operators.

Clearly, the inverse decay rates $\Gamma^{-1}$ is minimized at $\Lambda_{11}=1$, irrespective of decay modes and operator types.
Further imposing eq.~(\ref{eq:eps-lam_range}), 
$\Gamma^{-1}$ is minimized at $\eps_{q_3}=1$ and $\lambda_\yuk=4\pi$ for the $qqq\ell$ and $qque$ operators and 
at $\eps_{q_3}=1/\lambda_\yuk=1/(4\pi)$ for the $duue$ and $duq\ell$ operators. 
The normalization of $\eps_{q_3}$ and $\lambda_\yuk$ in eq.~(\ref{eq:Gamma_scaling}) is inspired by minimizing $\Gamma^{-1}$ for each BNV operator.

Varying $\eps_{q_3}\in[\frac{1}{4\pi},\,1]$ with $\Lambda_{11}=5$ and $\lambda_{\yuk}=4\pi$, 
the proton lifetimes for the decays into the first generation leptons are in the following range:
\begin{align}
\Gamma^{-1}(p \to \pi^0 e^+)
	& \simeq (7.7\,\mbox{--}\,6800) \times 10^{34}\,{\rm yrs} \nonumber\\
\Gamma^{-1}(p \to \pi^+ \overline\nu_e)
	& \simeq (4.7\,\mbox{--}\,2000) \times 10^{35}\,{\rm yrs} \nonumber\\
\Gamma^{-1}(p \to K^0 e^+)
	& \simeq (2.0\,\mbox{--}\,870) \times 10^{34}\,{\rm yrs} \label{eq:Gamma_PtoE_min}\\
\Gamma^{-1}(p \to K^+\overline\nu_e)
	& \simeq (1.1\,\mbox{--}\,520) \times 10^{34}\,{\rm yrs} \nonumber\\
\Gamma^{-1}(p \to \eta e^+)
	& \simeq (1.9\,\mbox{--}\,6200) \times 10^{35}\,{\rm yrs} \nonumber
\end{align}
Here we take $c_{qqq\ell}=c_{qque}=c_{duue}=c_{duq\ell}=1$. 
These are all above the current 90\% confidence-level (CL) limits from the Super-Kamiokande experiment,
\begin{align}
\Gamma^{-1}(p \to \pi^0 e^+)
	& \geq 2.4 \times 10^{34}\,{\rm yrs} ~\mbox{\cite{Super-Kamiokande:2020wjk}} \nonumber\\
\Gamma^{-1}(p \to \pi^+ \overline\nu)
	& \geq 3.9 \times 10^{32}\,{\rm yrs} ~\mbox{\cite{Super-Kamiokande:2013rwg}} \nonumber\\
\Gamma^{-1}(p \to K^0 e^+)
	& \geq 0.10 \times 10^{34}\,{\rm yrs} ~\mbox{\cite{Super-Kamiokande:2005lev}} \\
\Gamma^{-1}(p \to K^+\overline\nu)
	& \geq 0.59 \times 10^{34}\,{\rm yrs} ~\mbox{\cite{Super-Kamiokande:2014otb}} \nonumber\\
\Gamma^{-1}(p \to \eta e^+)
	& \geq 1.4 \times 10^{34}\,{\rm yrs} ~\mbox{\cite{Super-Kamiokande:2024qbv}} \nonumber
\end{align}
Similarly, for the decays into the second generation leptons, varying $\eps_{q_3}\in[\frac{1}{4\pi},\,1]$ with $\Lambda_{11}=5$ and $\lambda_{\yuk}=4\pi$ leads to the proton lifetimes in the following range:
\begin{align}
\Gamma^{-1}(p \to \pi^0 \mu^+)
	& \simeq (3.9\,\mbox{--}\,4200) \times 10^{31}\,{\rm yrs} \nonumber\\
\Gamma^{-1}(p \to \pi^+ \overline\nu_\mu) 
	& \simeq (2.1\,\mbox{--}\,940) \times 10^{34}\,{\rm yrs} \nonumber\\
\Gamma^{-1}(p \to K^0 \mu^+)
	& \simeq (3.9\,\mbox{--}\,470) \times 10^{31}\,{\rm yrs} \label{eq:Gamma_PtoM_min}\\
\Gamma^{-1}(p \to K^+ \overline\nu_\mu)
	& \simeq (5.2\,\mbox{--}\,2400) \times 10^{32}\,{\rm yrs} \nonumber\\
\Gamma^{-1}(p \to \eta \mu^+)
	& \simeq (0.94\,\mbox{--}\,19000) \times 10^{32}\,{\rm yrs} \nonumber
\end{align}
Interestingly, these lifetime ranges are partly covered by the current Super-Kamiokande limits,
\begin{align}
\Gamma^{-1}(p \to \pi^0 \mu^+)
	& \geq 1.6 \times 10^{34}\,{\rm yrs} ~\mbox{\cite{Super-Kamiokande:2020wjk}} \nonumber\\
\Gamma^{-1}(p \to \pi^+ \overline\nu)
	& \geq 3.9 \times 10^{32}\,{\rm yrs} ~\mbox{\cite{Super-Kamiokande:2013rwg}} \nonumber\\
\Gamma^{-1}(p \to K^0 \mu^+)
	& \geq 0.36 \times 10^{34}\,{\rm yrs} ~\mbox{\cite{Super-Kamiokande:2022egr}} \\
\Gamma^{-1}(p \to K^+\overline\nu)
	& \geq 0.59 \times 10^{34}\,{\rm yrs} ~\mbox{\cite{Super-Kamiokande:2014otb}} \nonumber\\
\Gamma^{-1}(p \to \eta \mu^+)
	& \geq 0.73 \times 10^{34}\,{\rm yrs} ~\mbox{\cite{Super-Kamiokande:2024qbv}} \nonumber
\end{align}
and Hyper-Kamiokande projected sensitivity, $\Gamma^{-1}(p \to \pi^0 \mu^+)_{\rm HK} \geq 7.7 \times 10^{34}\,{\rm yrs}$, can cover the remaining range \cite{Hyper-Kamiokande:2018ofw}.

\begin{figure}[t]
\centering
\includegraphics[width=0.48\textwidth]{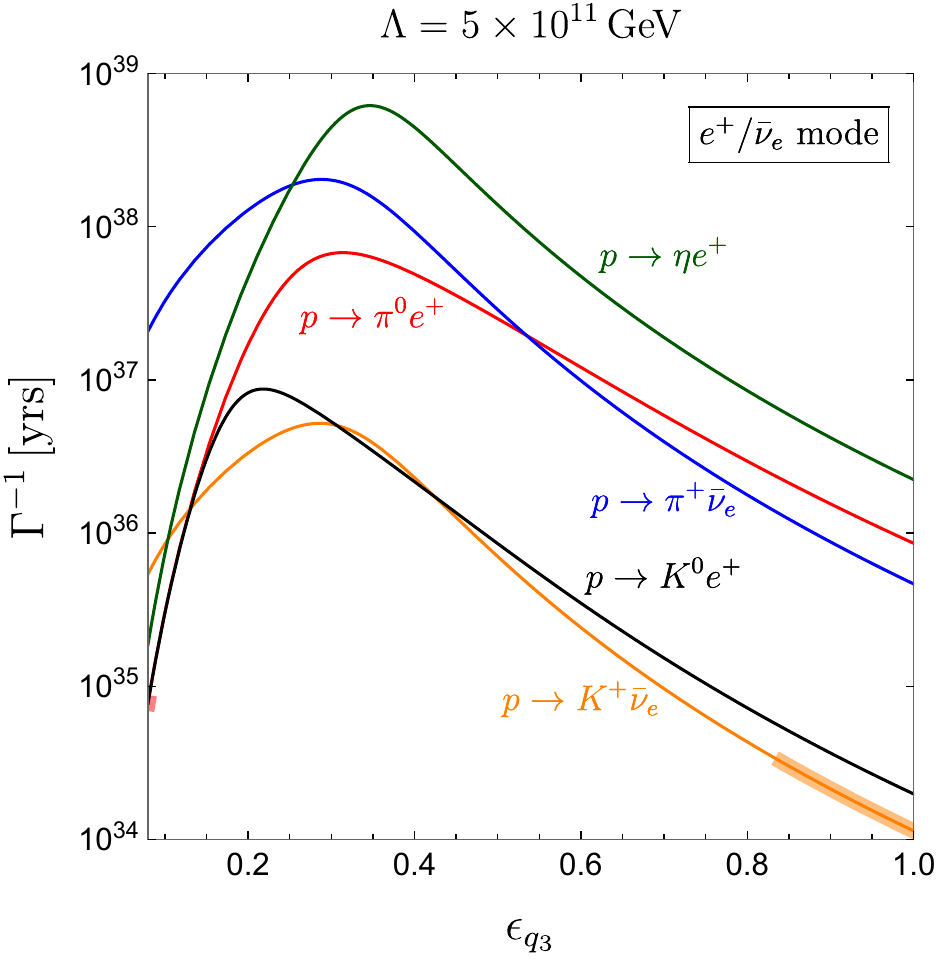}
\includegraphics[width=0.48\textwidth]{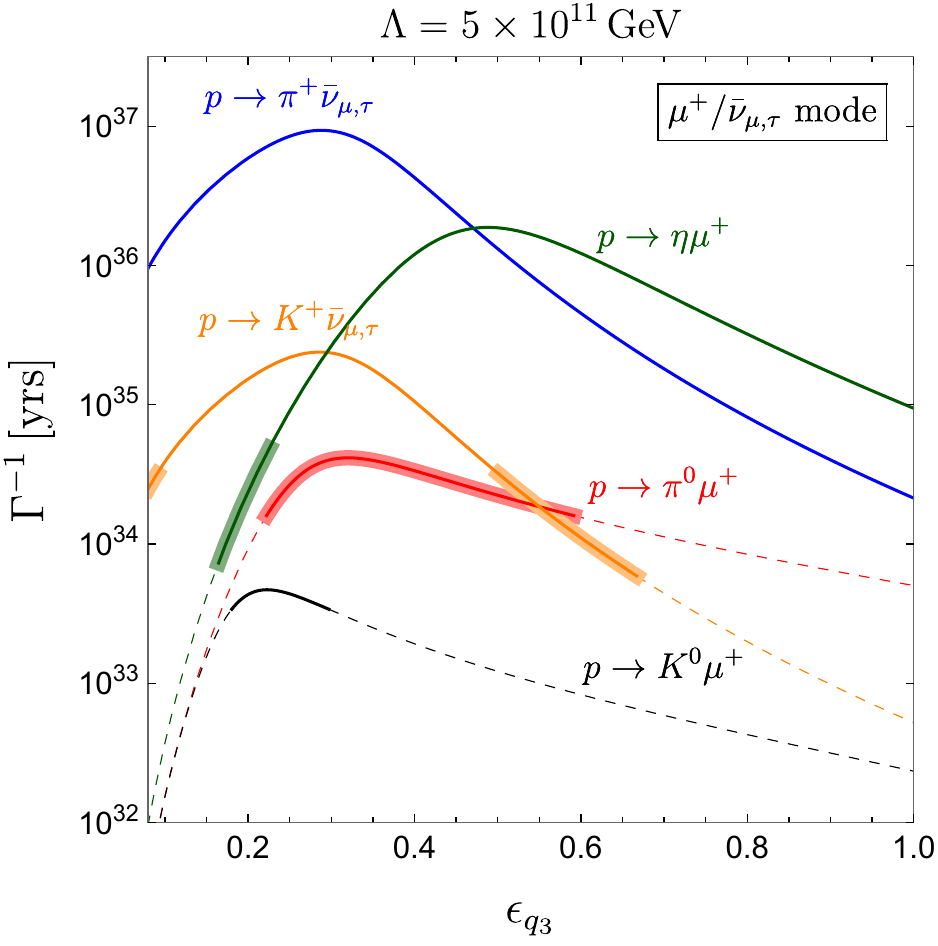}
\caption{
The $\epsilon_{q_3}$-dependence of the proton lifetime for each decay mode 
for $\Lambda=5\times10^{11}$~GeV with $\lambda_{\yuk}=4\pi$ and $c_{qqq\ell}=c_{qque}=c_{duue}=c_{duq\ell}=1$.
In each curve, 
the dashed part is excluded by the Super-Kamiokande experiment and 
the thick part is reached by the future Hyper-Kamiokande experiment.
}
\label{fig:lifetime}
\end{figure}

The $\epsilon_{q_3}$-dependence of the proton lifetime for each decay mode is shown in fig.~\ref{fig:lifetime} with $\Lambda=5\times10^{11}$~GeV, $\lambda_{\rm yuk}=4\pi$, and $c_{qqq\ell}=c_{qque}=c_{duue}=c_{duq\ell}=1$.
In each curve, 
the dashed part is excluded by the Super-Kamiokande experiment and 
the thick part is reached by the future Hyper-Kamiokande experiment.
It is observed that two decay modes, $p \to \pi^0 \mu^+$ and $p \to K^0 \mu^+$, provide the leading constraints and the current Super-Kamiokande limits \cite{Super-Kamiokande:2020wjk, Super-Kamiokande:2022egr} restrict the viable parameter range to be $\epsilon_{q_3} \simeq 0.22$\,--\,0.3. 
This range would be within the future Hyper-Kamiokande reach: 
$\Gamma^{-1}(p \to \pi^0 \mu^+)_{\rm HK} \geq 7.7 \times 10^{34}\,{\rm yrs}$ \cite{Hyper-Kamiokande:2018ofw}.
Note that these predictions should be regarded as order-of-magnitude estimates 
as there are unknown $O(1)$ coefficients in each higher-dimensional operator and Yukawa matrix.
Nonetheless, the estimates capture the overall parameter dependence and provide a rough target range for the lifetimes to be explored in future experiments.

\begin{figure}[t]
\centering
\adjustbox{valign=c}{\includegraphics[width=0.4\textwidth]{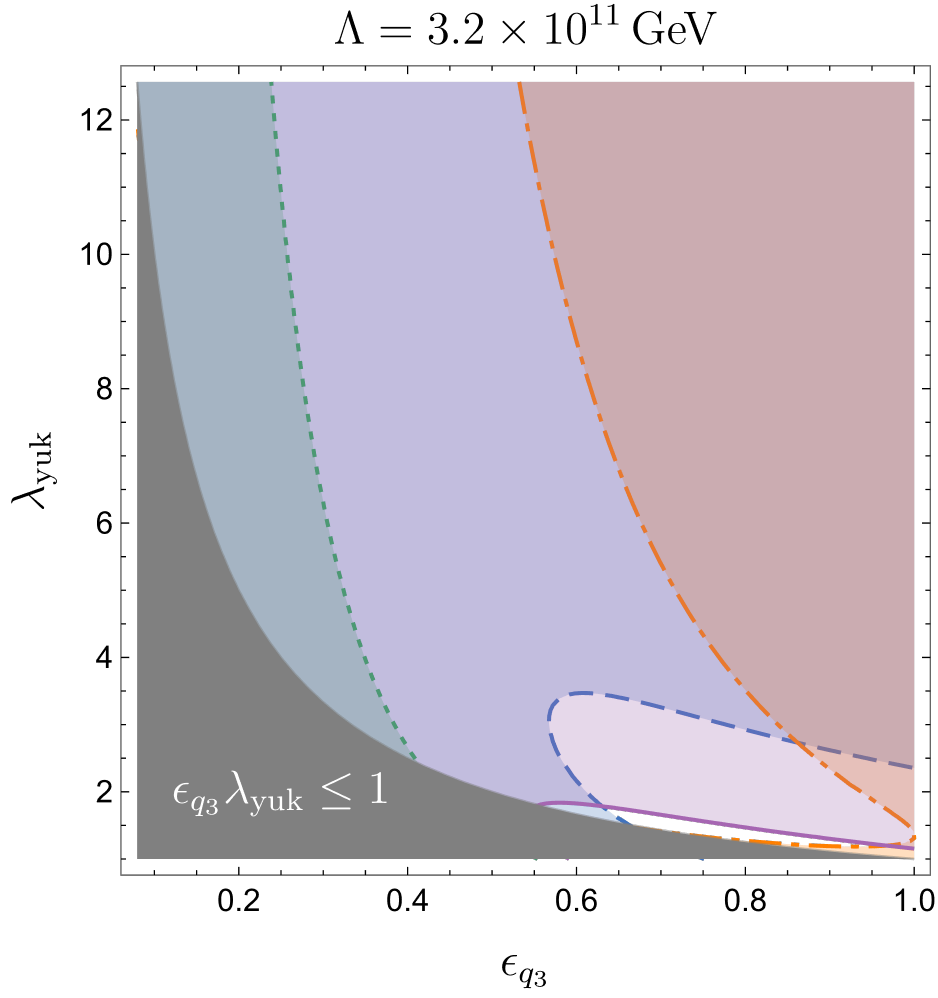}}
\adjustbox{valign=c}{\includegraphics[width=0.4\textwidth]{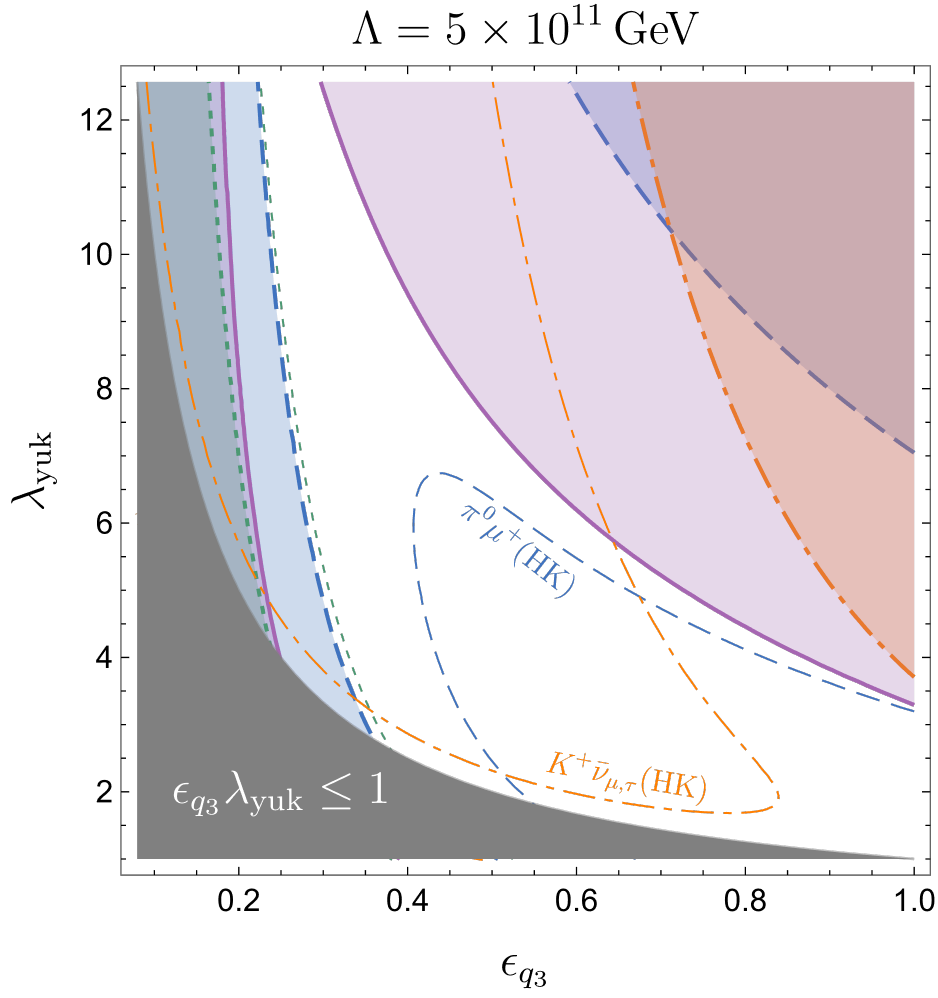}}

\vspace{0.2cm}

\adjustbox{valign=c}{\includegraphics[width=0.4\textwidth]{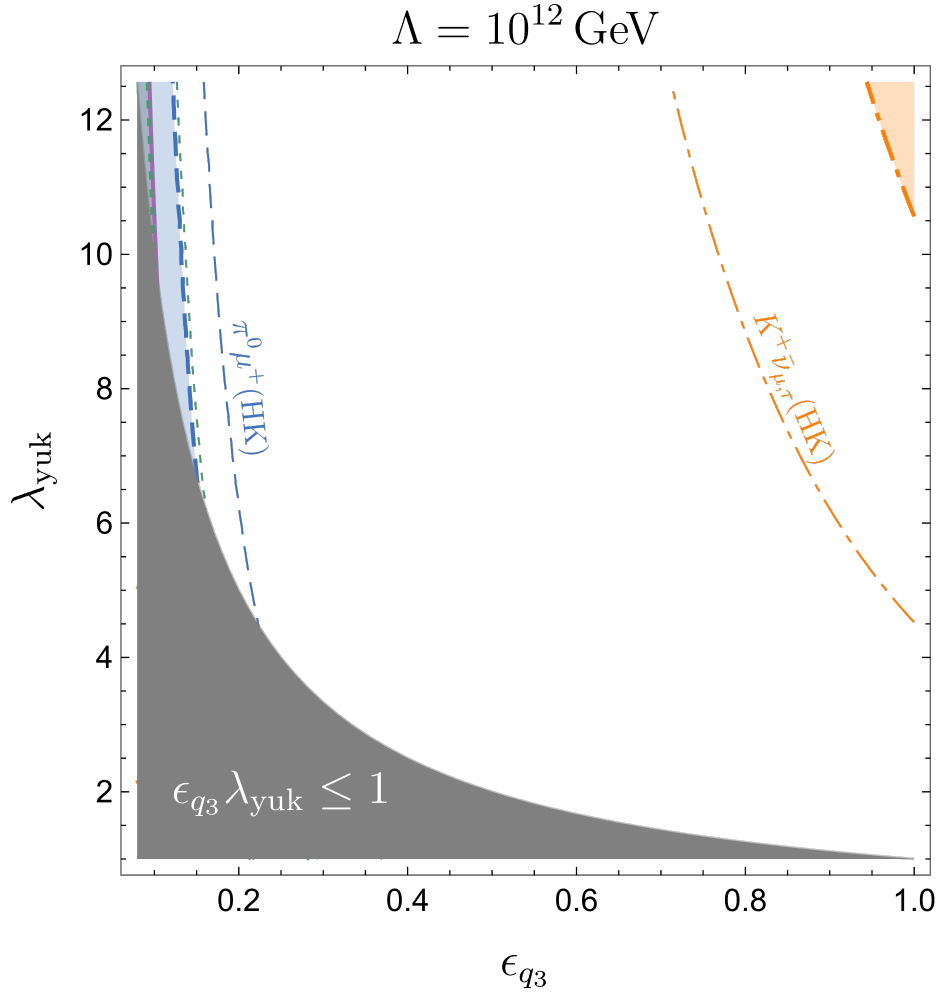}}
\hspace{1cm}
\adjustbox{valign=c}{\includegraphics[width=0.18\textwidth]{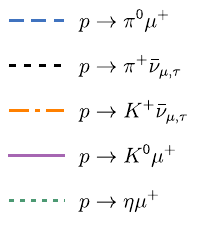}}
\hspace{2cm}
\caption{
The current proton decay bounds and future Hyper-Kamiokande sensitivities in the ($\epsilon_{q_3}$, $\lambda_{\rm yuk})$ plane with $c_{qqq\ell}=c_{qque}=c_{duue}=c_{duq\ell}=1$ for $\Lambda=3.2\times10^{11}\,\GeV$ (top-left), $5\times10^{11}\,\GeV$ (top-right), and 
$10^{12}$~GeV (bottom-left).
The shaded regions are excluded by the Super-Kamiokande experiment \cite{Super-Kamiokande:2020wjk, Super-Kamiokande:2022egr, Super-Kamiokande:2024qbv}.
In the gray region, one finds $\epsilon_{q_3}\lambda_{\rm yuk} \leq 1$ which is inconsistent with the top yukawa coupling, $\lambda_{\rm yuk}\epsilon_{q_3}\epsilon_{u_3}=y_t\simeq1$.
}
\label{fig:eps-lam}
\end{figure}

In fig.~\ref{fig:eps-lam}, the current proton decay bounds are summarized in the $(\epsilon_{q_3},\,\lambda_{\rm yuk})$ plane for $\Lambda=3.2\times10^{11}$~GeV (top-left), $5\times10^{11}$~GeV (top-right), and $10^{12}$~GeV (bottom-left), 
together with future Hyper-Kamiokande sensitivities \cite{Hyper-Kamiokande:2018ofw}. 
The colored shaded regions are excluded by the Super-Kamiokande \cite{Super-Kamiokande:2020wjk, Super-Kamiokande:2022egr, Super-Kamiokande:2024qbv}.
In the gray region on the bottom left, one finds $\epsilon_{q_3}\lambda_{\rm yuk} \leq 1$ which is inconsistent with the top Yukawa coupling, $\lambda_{\rm yuk}\epsilon_{q_3}\epsilon_{u_3}=y_t\simeq1$.
Three decay modes, $p \to \pi^0 \mu^+$, $p \to K^0 \mu^+$, and $p \to K^+ \bar{\nu}_{\mu,\tau}$, provide the leading constraints, limiting the parameter spaces complementarily.
A narrow yet viable parameter region can be found for $\Lambda=3.2\times10^{11}$~GeV, indicating that $\Lambda\sim3\times10^{11}$~GeV is the lowest scale allowed by the proton decay search.

It is worth mentioning that the latest Super-Kamiokande search has observed one candidate event for $p\to\pi^0\mu^+$, but not for $p\to\pi^0e^+$ \cite{Super-Kamiokande:2020wjk}. 
If this event is due to the proton decay signal, the future Hyper-Kamiokande search will observe more events for $p\to\pi^0\mu^+$. 
Furthermore, if null observation of $p \to \pi^0 e^+$ persists in that case, 
it would indicate the new physics structure characterized by the $\epsilon$ scheme and the scale $\Lambda={\cal O}(10^{11})$~GeV. 
To confirm such a new physics structure, it would also be important to search for $p\to K^0 \mu^+$ and $p\to K^+ \nu$ in future experiments.

\begin{figure}[t]
\centering
\includegraphics[width=0.65\textwidth]{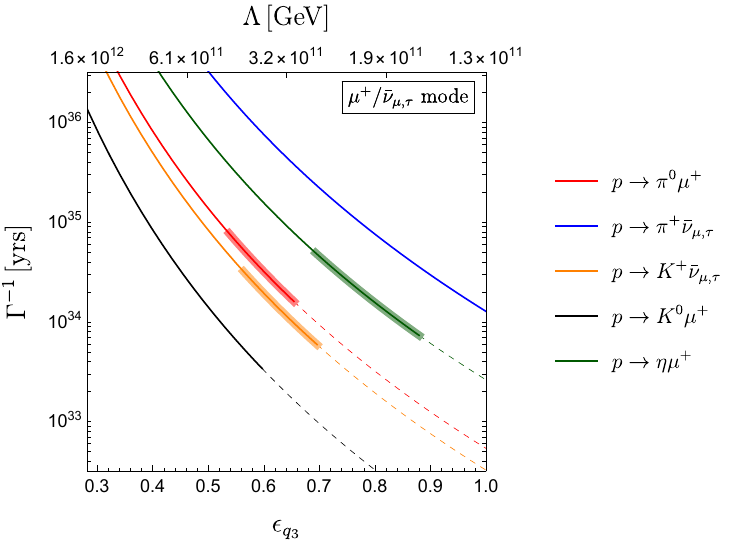}
\caption{
The $\epsilon_{q_3}$-dependence of the proton lifetime for each decay mode with the GUT-inspired relations and $c_{qqq\ell}=c_{qque}=c_{duue}=c_{duq\ell}=1$.
The remaining two parameters are determined by $\lambda_{\rm yuk} \sim (\epsilon_{q_3})^{-2}$ and $\Lambda \sim 1.3 \times 10^{11}\,{\rm GeV} \times (\epsilon_{q_3})^{-2}$ with $\epsilon_{q_3}\in[\frac{1}{\sqrt{4\pi}},\,1]$.
In each curve, 
the dashed part is excluded by the Super-Kamiokande experiment and 
the thick part is reached by the future Hyper-Kamiokande experiment.
}
\label{fig:lifetime_GUT}
\end{figure}

Lastly, we study the implication from the GUT-inspired relations, $\epsilon_q\sim\epsilon_u\sim\epsilon_e$ and $\epsilon_d\sim\epsilon_\ell$.
With these two relations, we only have one free parameter.
Figure \ref{fig:lifetime_GUT} shows the same plot as fig.~\ref{fig:lifetime}, but with $\epsilon_{q_3}$ chosen to be the free parameter, and the other two parameters, $\lambda_{\rm yuk}$ and $\Lambda$, determined by $\lambda_{\rm yuk} \sim (\epsilon_{q_3})^{-2}$ and $\Lambda \sim 1.3 \times 10^{11}\,{\rm GeV} \times (\epsilon_{q_3})^{-2}$, with $\epsilon_{q_3}\in[\frac{1}{\sqrt{4\pi}}\,,1]$.
In each curve, 
the dashed part is excluded by the Super-Kamiokande experiment and 
the thick part is reached by the future Hyper-Kamiokande experiment.
One can see that currently $p \to K^0 \mu^+$ provides the strongest constraint, restricting the allowed range to be $\epsilon_{q_3} \lesssim 0.6$, which corresponds to $\Lambda\gtrsim4\times10^{11}$~GeV.

\section{Summary}
\label{sec:summary}

After the discovery of the Higgs boson as a light scalar, 
the scale of new physics is the central question of particle physics.
If we ignore the naturalness, the most extreme scenario would be
that the next scale $\Lambda$ is much higher than the electroweak scale and the Standard Model
is replaced by a more fundamental theory such as quantum gravity
above the scale $\Lambda$. 
One of the hints to guess the scale $\Lambda$ is the neutrino masses
that can be explained easily by the dimension-five operator, $\ell\ell HH/\Lambda$. 
The scale which appears in the denominator gives a first guess for the scale $\Lambda$.
Taking this picture, and considering the flavor structure in the Yukawa matrices,
we found that the scale $\Lambda \sim 10^{11}$~GeV fits well with all the flavor observables.
In that case, one can also expect other higher dimensional operators with the suppression of $1/\Lambda^n$ to appear below the scale $\Lambda$ and leave observable effects at low energies, which serve as an important consistency test on this scenario. 
As examples, we discussed the proton decays induced by general dimension-six BNV operators with the flavor structure expected from the Yukawa hierarchies, and found that the proton lifetimes for all decay modes are consistent with the current Super-Kamiokande limits barely for $\Lambda \sim 10^{11}$~GeV for decay modes with the muon or the Kaon in the final states. 
Interestingly, some decay modes, such as $p\to\pi^0\mu^+$, can have a lifetime of order of $10^{34}\,{\rm yrs}$, which could be potentially observed at the future Hyper-Kamiokande experiment.

The discussion points to the importance of the decay mode $p \to \pi^0 \mu^+$, that has
actually been observed already at the Super-Kamiokande experiment although
the event is consistent with the expected background events~\cite{Super-Kamiokande:2020wjk}. 
This mode will be particularly interesting for the future search at Hyper-Kamiokande.
The significance of the $\mu^+$ mode seems
to be more general when considering the flavor structure in the proton decays, 
for example, see refs.~\cite{Ibe:2024cbt,Ibe:2024cvi,Chitose:2025bvl,Beneito:2025fzf}.
It is noteworthy that the nature of fermion hierarchy
can be learned by looking at the proton decays in future.

The physics of the scale $\Lambda$ should also explain mysteries in the Standard Model
such as dark matter and baryon asymmetry of the Universe. 
One possibility is the presence of an axion associated with the decay constant of order of $\Lambda$,
with an initial condition in which the axion has a non-zero velocity after inflation~\cite{Co:2019wyp,Co:2019jts,Co:2020dya,Domcke:2020kcp}.
This represents a particularly intriguing scenario, and model-building of the physics above the scale $\Lambda$
will be a quite interesting future direction.

\section*{Acknowledgements}
This work is supported in part by JSPS KAKENHI Grant Numbers
JP22K21350 (R.K. and S.O.) and JP25K17401 (S.O.).
The work of S.O. is also supported by an appointment to the JRG Program at the APCTP through the Science and Technology Promotion Fund and Lottery Fund of the Korean Government and by the Korean Local Governments -- Gyeongsangbuk-do Province and Pohang City.

\appendix

\section{Matching SMEFT with LEFT}

When computing the nucleon lifetimes, one has to match SMEFT with LEFT at the electroweak scale. 
The LEFT Lagrangian is given by 
\beq
\L_{\rm LEFT} = \sum_I \sum_{i,j,k,l} L_{I,\,ijkl} \, {\cal O}_{I,\,ijkl}^{\rm LEFT} \,,
\eeq
where the BNV LEFT operators are listed in Table \ref{tab:fig:LEFT-BNV}. 
Tree-level matching conditions between the SMEFT and LEFT coefficients are given by \cite{Jenkins:2017jig}
\begin{align}
\(L_{udd}^{S,LL}\)_{ijkl} 
	& = V_{j'j} V_{k'k} \(C_{qqql,\,k'j'il} - C_{qqql,\,j'k'il} + C_{qqql,\,j'ik'l} \) \\
\(L_{duu}^{S,LL}\)_{ijkl} 
	& = V_{i'i} \(C_{qqql,\,kji'l} - C_{qqql,\,jki'l} + C_{qqql,\,ji'kl} \) \\
\(L_{duu}^{S,LR}\)_{ijkl} 
	& = -V_{i'i} \(C_{qque,\,i'jkl}  + C_{qque,\,ji'kl} \) \\
\(L_{duu}^{S,RL}\)_{ijkl} 
	& = C_{duql,\,ijkl} \\
\(L_{dud}^{S,RL}\)_{ijkl} 
	& = -V_{k'k} \, C_{duql,\,ijk'l} \label{eq:LSRL_dud}\\
\(L_{duu}^{S,RR}\)_{ijkl} 
	& = C_{duue,\,ijkl} 
\end{align}
where $V_{ij}$ denotes the CKM matrix element and we take the up-quark basis in the SMEFT. 

Note that our parameterization, introduced in eq.~(\ref{eq:eps_qi}), leads to an artificial cancellation, once combined with the above matching equations. 
To be more concrete, eq.~(\ref{eq:LSRL_dud}) results in $\left(L_{dud}^{S,RL}\right)_{ijkl} \propto -V_{k'k} V_{k'3} \, C_{duql,\,ij3l} \simeq -\delta_{k3} \, C_{duql,\,ij3l} \simeq 0$ for $k=1,2$, thereby significantly suppressing proton decays. 
For a more general parametrization $\epsilon_{q_i} = \epsilon_{q_3} \frac{c_i}{c_3} \frac{|V_{i3}|}{|V_{33}|}$ with $c_i$ being random numbers of ${\cal O}(1)$, such cancellation can only occur accidentally and unphysically when all $c_i$ have precisely the same values.
To avoid this artificial cancellation while maintaining the flavor structure we consider in this paper, we approximate $V_{ij} = \delta_{ij}$ only in the SMEFT-LEFT matching equations in our analysis.

\renewcommand{\arraystretch}{1.3}
\begin{table}[t!]
    \centering
    \begin{tabular}{c|c||c|c}
    \hline
        ${\cal O}_{udd}^{S,LL}$ & $\eps_{abc} (u_{Li}^{a T} C d_{Lj}^b) (d_{Lk}^{c T} C \nu_{Ll})$ & ${\cal O}_{uud}^{S,RL}$ & $\eps_{abc} (u_{Ri}^{a T} C u_{Rj}^b) (d_{Lk}^{c T} C e_{Ll})$ \\
        ${\cal O}_{duu}^{S,LL}$ & $\eps_{abc} (d_{Li}^{a T} C u_{Lj}^b) (u_{Lk}^{c T} C e_{Ll})$ & ${\cal O}_{duu}^{S,RL}$ & $\eps_{abc} (d_{Ri}^{a T} C u_{Rj}^b) (u_{Lk}^{c T} C e_{Ll})$ \\
        ${\cal O}_{uud}^{S,LR}$ & $\eps_{abc} (u_{Li}^{a T} C u_{Lj}^b) (d_{Rk}^{c T} C e_{Rl})$ & ${\cal O}_{dud}^{S,RL}$ & $\eps_{abc} (d_{Ri}^{a T} C u_{Rj}^b) (d_{Lk}^{c T} C \nu_{Ll})$ \\
        ${\cal O}_{duu}^{S,LR}$ & $\eps_{abc} (d_{Li}^{a T} C u_{Lj}^b) (u_{Rk}^{c T} C e_{Rl})$ & ${\cal O}_{ddu}^{S,RL}$ & $\eps_{abc} (d_{Ri}^{a T} C d_{Rj}^b) (u_{Lk}^{c T} C \nu_{Ll})$ \\
        & & ${\cal O}_{duu}^{S,RR}$ & $\eps_{abc} (d_{Ri}^{a T} C u_{Rj}^\beta) (u_{Rk}^{c T} C e_{Rl})$ \\
    \hline
    \end{tabular}
    \caption{
    The basis of the $\Delta B=\Delta L=1$ LEFT operators \cite{Jenkins:2017jig}. 
    Here, $a,b,c=1,2,3$ denote the color indices and $i,j,k,l=1,2,3$ the flavor indices in the mass basis.
    }
    \label{tab:fig:LEFT-BNV}
\end{table}
\renewcommand{\arraystretch}{1}

\section{Analytical expressions of nucleon decay rates}

Some details of the nucleon decay calculation are provided here.

\subsection{Hadronic inputs}

Due to parity conservation of strong interactions and isospin symmetry, there are only two independent nuclear matrix elements for nucleon decays. 
These are parameterized in terms of two real parameters $\alpha$ and $\beta$ \cite{JLQCD:1999dld},
\beq
\langle 0 | \epsilon^{abc} (\bar{u}_a^\dag \bar{d}_b^\dag) u_c | p^{(s)} \rangle = \alpha P_L u_p^{(s)} \,,\quad
\langle 0 | \epsilon^{abc} (u_a d_b) u_c | p^{(s)} \rangle = \beta P_L u_p^{(s)} \,,
\label{eq:nmatrix_p}
\eeq
where $a,b,c=1,2,3$ denote the color indices, $s$ the proton spin, $u_p^{(s)}$ the four-component proton spinor and $P_{L,R}$ the chirality projection. 
The matrix elements for neutron (with its four-component spinor $u_n^{(s)}$) are obtained from isospin symmetry:
\beq
\langle 0 | \epsilon^{abc} (\bar{d}_a^\dag \bar{u}_b^\dag) d_c | n^{(s)} \rangle = \alpha P_L u_n^{(s)} \,,\quad
\langle 0 | \epsilon^{abc} (d_a u_b) d_c | n^{(s)} \rangle = \beta P_L u_n^{(s)} \,.
\label{eq:nmatrix_n}
\eeq
The ones for right-handed spinors are related to eqs.~(\ref{eq:nmatrix_p})-(\ref{eq:nmatrix_n}) through parity conservation:
\begin{align}
\langle 0 | \epsilon^{abc} (u_a d_b) \bar{u}_c^\dag | p^{(s)} \rangle & = - \alpha P_R u_p^{(s)} \,,\quad
\langle 0 | \epsilon^{abc} (\bar{u}_a^\dag \bar{d}_b^\dag) \bar{u}_c^\dag | p^{(s)} \rangle = - \beta P_R u_p^{(s)} \,,\\
\langle 0 | \epsilon^{abc} (d_a u_b) \bar{d}_c^\dag | n^{(s)} \rangle & = - \alpha P_R u_n^{(s)} \,,\quad
\langle 0 | \epsilon^{abc} (\bar{d}_a^\dag \bar{u}_b^\dag) \bar{d}_c^\dag | n^{(s)} \rangle = - \beta P_R u_n^{(s)} \,.
\end{align}
The values of $\alpha$ and $\beta$ are determined by lattice QCD calculation~\cite{Yoo:2021gql}
\begin{align}
\alpha = -0.01257(111)\,\GeV^3 \,,\quad
\beta = 0.01269(107)\,\GeV^3\,.
\end{align}

Contributions from hyperon pole diagrams are calculated using BChPT. 
The leading order Lagrangian for the relevant baryon-meson interactions is given by 
\beq
\L_0 = \frac{1}{8} f_\pi^2 \, {\rm tr}(\partial_\mu\Sigma \partial^\mu\Sigma^\dag) 
	+ {\rm Tr} \( \bar{B}(i\slashed{D}-M_B)B \) 
	- \frac{D}{2} {\rm Tr} \( \bar{B}\gamma^\mu\gamma_5 \{\xi_\mu, B\} \)
	- \frac{F}{2} {\rm Tr} \( \bar{B}\gamma^\mu\gamma_5 [\xi_\mu, B] \) 
\eeq
where $f_\pi=130\,\MeV$ is the pion decay constant. 
Here, $B$ denotes the octet baryon field, 
\beq
B = \begin{pmatrix} 
    \frac{\Sigma^0}{\sqrt{2}}+\frac{\Lambda^0}{\sqrt{6}} & \Sigma^+ & p \\
    \Sigma^- & -\frac{\Sigma^0}{\sqrt{2}}+\frac{\Lambda^0}{\sqrt{6}} & n \\
    \Xi^- & \Xi^0 & -\sqrt{\frac{2}{3}}\Lambda^0
    \end{pmatrix}
\eeq
and the octet meson field $M$ is encoded in $\xi=e^{iM/(2f_\pi)}$:
\beq
M = \begin{pmatrix} 
    \frac{\pi^0}{\sqrt{2}}+\frac{\eta}{\sqrt{6}} & \pi^+ & K^+ \\
    \pi^- & -\frac{\pi^0}{\sqrt{2}}+\frac{\eta}{\sqrt{6}} & K^0 \\
    K^- & \bar{K}^0 & -\sqrt{\frac{2}{3}}\eta
    \end{pmatrix}
\eeq
Note $\Sigma=\xi^2=e^{iM/f_\pi}$ and $\xi_\mu=i\left[\xi^\dagger\partial_\mu\xi-\xi\partial_\mu\xi^\dagger\right]$.
The low-energy constants $D$ and $F$ are obtained from a phenomenological analysis \cite{Aoki:2008ku}, 
\beq
D = 0.80(1) \,,\quad
F=0.47(1) \,,
\eeq
or calculated by lattice simulation \cite{Bali:2022qja}
\beq
D = 0.730(11) \,,\quad
F = 0.447^{(6)}_{(7)} \,.
\eeq
We use the latter set of the low-energy constants in the analysis.

\subsection{Decay rates with $|\Delta(B-L)|=0$}

Combining the LEFT Lagrangian with the hadronic inputs, 
one can obtain the nucleon decay rates with $|\Delta(B-L)|=0$ (see also Appendix F of \cite{Nath:2006ut}), 
\begin{align}
\Gamma(p \to \pi^0 e^+_i) 
	&=\frac{(m_p^2-m_{\pi^0}^2)^2}{32\pi f_\pi^2 m_p^3}  \frac{(1+D+F)^2}{2} \left(\left|\alpha\(L_{duu}^{S,RL}\)_{111i} +\beta\(L_{duu}^{SLL}\)_{111i}\right|^2\right. \nonumber\\
	&\quad\left.{} +\left|\alpha\(L_{duu}^{S,LR}\)_{111i}+\beta\(L_{duu}^{S,RR}\)_{111i}\right|^2\right) \,,\label{eq:Gamma_PtoPiE}\\
\Gamma(n \to \pi^- e^+_i) 
	&=\frac{(m_n^2-m_{\pi^-}^2)^2}{32\pi f_\pi^2 m_n^3} \frac{3(1+D+F)^2}{2} \left(\left|\alpha\(L_{duu}^{S,RL}\)_{111i}+\beta\(L_{duu}^{SLL}\)_{111i} \right|^2\right. \nonumber\\
	&\quad\left.{} +\left|\alpha \(L_{duu}^{S,LR}\)_{111i} + \beta \(L_{duu}^{S,RR}\)_{111i} \right|^2 \) \,,\\
\Gamma(p \to \pi^+ \bar{\nu}_i) 
	& = \frac{(m_p^2-m_{\pi^+}^2)^2}{32\pi f_\pi^2 m_p^3} (1+D+F)^2 \left| \alpha \(L_{dud}^{S,RL}\)_{111i} + \beta \(L_{udd}^{S,LL}\)_{111i} \right|^2 \,,\\
\Gamma(n \to \pi^0 \bar{\nu}_i) 
	& = \frac{(m_n^2-m_{\pi^0}^2)^2}{32\pi f_\pi^2 m_n^3} \frac{(1+D+F)^2}{2} \left| \alpha \(L_{dud}^{S,RL}\)_{111i} + \beta \(L_{udd}^{S,LL}\)_{111i} \right|^2 \,.
\end{align}
It is easy to see that there are correlations between nucleon decay modes, 
\begin{align}
\Gamma(p \to \pi^0 e^+_i)  & \simeq \frac{1}{3} \Gamma(n \to \pi^- e^+_i) \,,\\
\Gamma(p \to \pi^+ \bar{\nu}_i) & \simeq 2 \Gamma(n \to \pi^0 \bar{\nu}_i) \,.
\end{align}
The rates for other decay modes are 
\begin{align}
\Gamma(p \to K^+ \bar{\nu}_i) 
	& = \frac{(m_p^2-m_K^2)^2}{32\pi f_\pi^2 m_p^3}  
		\left| \left\{ \beta \(L_{udd}^{S,LL}\)_{112i} + \alpha \(L_{dud}^{S,RL}\)_{112i} \right\} \(1+\frac{m_p}{3m_\Lambda}(D+3F)\) \right. \nonumber\\
	& \quad \left.{}
			+ \left\{ \beta \(L_{udd}^{S,LL}\)_{121i} + \alpha \(L_{dud}^{S,RL}\)_{211i} \right\} \(\frac{m_p}{2m_{\Sigma^0}}(D-F)+\frac{m_p}{6m_\Lambda}(D+3F)\) \right|^2 \,,\\
\Gamma(n \to K^0 \bar{\nu}_i) 
	& = \frac{(m_n^2-m_K^2)^2}{32\pi f_\pi^2 m_n^3} 
			\left| \alpha \(L_{dud}^{S,RL}\)_{211i} \(-1-\frac{m_n}{2m_{\Sigma^0}}(D-F)+\frac{m_n}{6m_\Lambda}(D+3F)\) \right. \nonumber\\
	& \quad	+ \beta \(L_{udd}^{S,LL}\)_{121i} \(1-\frac{m_n}{2m_{\Sigma^0}}(D-F)+\frac{m_n}{6m_\Lambda}(D+3F)\) \nonumber\\
	& \quad \left.{}
			+ \left\{ \alpha \(L_{dud}^{S,RL}\)_{112i} + \beta \(L_{udd}^{S,LL}\)_{112i} \right\} \(1+\frac{m_n}{3m_\Lambda}(D+3F)\) \right|^2 \,.\\
\Gamma(p \to K^0 e^+_i) 
	& = \frac{(m_p^2-m_K^2)^2}{32\pi f_\pi^2 m_p^3} \times \nonumber\\
	& \quad 	\left\{ \left| \alpha \(L_{duu}^{S,RL}\)_{211i} \(-1+\frac{m_p}{m_{\Sigma^0}}(D-F)\) 
			+ \beta \(L_{duu}^{S,LL}\)_{211i} \(1+\frac{m_p}{m_{\Sigma^0}}(D-F)\) \right|^2 \right. \nonumber\\
	& \quad \left.{}
			+ \left| \alpha \(L_{duu}^{S,LR}\)_{211i} \(-1+\frac{m_p}{m_{\Sigma^0}}(D-F)\) 
			+ \beta \(L_{duu}^{S,RR}\)_{211i} \(1+\frac{m_p}{m_{\Sigma^0}}(D-F)\) \right|^2 \right\} \,,\\            
\Gamma(p \to \eta^0 e^+_i) 
	& = \frac{(m_p^2-m_{\eta^0}^2)^2}{32\pi f_\pi^2 m_p^3} \frac{3}{2} 
		\left\{ \left| \alpha \(L_{duu}^{S,RL}\)_{111i} \(-\frac{1}{3}-\frac{D}{3}+F\) + \beta \(L_{duu}^{S,LL}\)_{111i} \(1-\frac{D}{3}+F\) \right|^2 \right.\nonumber\\
	& \quad \left.{}	+ \left|  \alpha \(L_{duu}^{S,LR}\)_{111i} \(-\frac{1}{3}-\frac{D}{3}+F\) + \beta \(L_{duu}^{S,RR}\)_{111i} \(1-\frac{D}{3}+F\) \right|^2 \right\} \,,\\
\Gamma(n \to \eta^0 \bar{\nu}_i) 
	& = \frac{(m_n^2-m_{\eta^0}^2)^2}{32\pi f_\pi^2 m_n^3} \frac{3}{2} 
		\left| \alpha \(L_{dud}^{S,RL}\)_{111i} \(-\frac{1}{3}-\frac{D}{3}+F\) + \beta \(L_{udd}^{S,LL}\)_{111i} \(1-\frac{D}{3}+F\) \right|^2 \,,
\end{align}
where the final state lepton masses are neglected.

\section{Details of $\Gamma(p \to \pi^0\mu^+)$ calculation}

Here, we provide the workflow to obtain the proton lifetimes, in order for readers to easily follow our calculation procedure.
As an example, we take the $p \to \pi^0\mu^+$ decay and $\lambda_{\rm BNV}=\lambda_{\rm yuk}$.

Given eq.~(\ref{eq:Gamma_PtoPiE}), we only need the SMEFT coefficients with the $(1112)$ flavor structure:
\begin{align}
C_{qqql,\,1112} & = 0.0287\,\lambda_{\rm yuk}^{1} \frac{c_{qqq\ell}}{\Lambda^2} \left(\frac{|V_{13}|^3}{|V_{33}|^3}\frac{|U_{23}|}{|U_{33}|}\right) \, \eps_{q_3}^3 \,\sqrt{\frac{\Lambda}{10^{12}\,\GeV}} \,,\\
C_{qque,\,1112} & = 0.352\,\lambda_{\rm yuk}^{1} \frac{c_{qque}}{\Lambda^2} \left(\frac{m_u}{m_t}\frac{|V_{13}|}{|V_{33}|}\frac{m_\mu}{m_\tau}\frac{|U_{33}|}{|U_{23}|}\right) \,\eps_{q_3}\,\sqrt{\frac{10^{12}\,\GeV}{\Lambda}} \,,\\
C_{duue,\,1112} & = 0.00834\,\lambda_{\rm yuk}^{-1} \frac{c_{duue}}{\Lambda^2} \left(\frac{m_u^2}{m_t^2}\frac{m_d}{m_b}\frac{|V_{33}|^3}{|V_{13}|^3}\frac{m_\mu}{m_\tau}\frac{|U_{33}|}{|U_{23}|}\right) \, \eps_{q_3}^{-3} \, \sqrt{\frac{10^{12}\,\GeV}{\Lambda}} \,,\\
C_{duql,\,1112} & = 6.81\times10^{-4}\,\lambda_{\rm yuk}^{-1} \frac{c_{duq\ell}}{\Lambda^2} \left(\frac{m_u}{m_t}\frac{m_d}{m_b}\frac{|V_{33}|}{|V_{13}|}\frac{|U_{23}|}{|U_{33}|}\right) \,\eps_{q_3}^{-1}\, \sqrt{\frac{\Lambda}{10^{12}\,\GeV}} \,. 
\end{align}
The SMEFT-LEFT matching equations (with $V_{ij}=\delta_{ij}$) lead to 
\begin{align}
\(L_{udd}^{S,LL}\)_{1112} & = C_{qqql,\,1112} \\
\(L_{duu}^{S,LL}\)_{1112} & = C_{qqql,\,1112} \\
\(L_{duu}^{S,LR}\)_{1112} & = -2C_{qque,\,1112} \\
\(L_{duu}^{S,RL}\)_{1112} & = C_{duql,\,1112} \\
\(L_{dud}^{S,RL}\)_{1112} & = -C_{duql,\,1112} \\
\(L_{duu}^{S,RR}\)_{1112} & = C_{duue,\,1112} 
\end{align}
Plugging these LEFT coefficients into eq.~(\ref{eq:Gamma_PtoPiE}), we end up with 
\begin{align}
\Gamma^{-1}(p \to \pi^0 \mu^+) 
&\sim 9.8\times10^{32}\,{\rm yrs}\,\frac{\Lambda_{11}^3}{c_{qqq\ell}^2 \epsilon_{q_3}^6} \left(\frac{4\pi}{\lambda_{\rm yuk}}\right)^2
+ 5.3\times10^{30}\,{\rm yrs}\,\frac{\Lambda_{11}^5}{c_{qque}^2 \epsilon_{q_3}^2} \left(\frac{4\pi}{\lambda_{\rm yuk}}\right)^2 \nonumber\\
& + 3.6\times10^{28}\,{\rm yrs}\,\frac{\Lambda_{11}^5}{c_{duue}^2} \left(\frac{\epsilon_{q_3}}{1/4\pi}\right)^6 \left(\frac{\lambda_{\rm yuk}}{4\pi}\right)^2
+ 4.4\times10^{34}\,{\rm yrs}\,\frac{\Lambda_{11}^3}{c_{duq\ell}^2} \left(\frac{\epsilon_{q_3}}{1/4\pi}\right)^2 \left(\frac{\lambda_{\rm yuk}}{4\pi}\right)^2
\end{align}
where we ignored the interference between the different LEFT operators and used the following set of input parameters: 
\begin{align}
m_u&=2.16\,\MeV,\quad
m_d=4.7\,\MeV,\quad
m_b=4.18\,\GeV,\quad
m_t=172\,\GeV,\quad\\
m_\mu&=106\,\MeV,\quad
m_\tau=1.78\,\GeV,\\
m_p&=938.27\,\MeV,\quad
m_{\pi^0}=134.98\,\MeV,\quad
f_\pi=130.41\,\MeV,\\
\alpha&=-0.01257\,\GeV^3,\quad
\beta=0.01269\,\GeV^3,\quad
D=0.730,\quad
F=0.447\\
|V_{13}|&=0.00363999,\quad
|V_{33}|=0.999108,\quad
|U_{23}|=0.7,\quad
|U_{33}|=0.7
\end{align}


\bibliographystyle{JHEP} 
\bibliography{bibliography}

@article{Dreiner:2008tw,
    author = "Dreiner, Herbi K. and Haber, Howard E. and Martin, Stephen P.",
    title = "{Two-component spinor techniques and Feynman rules for quantum field theory and supersymmetry}",
    eprint = "0812.1594",
    archivePrefix = "arXiv",
    primaryClass = "hep-ph",
    reportNumber = "BN-TH-2008-12, SCIPP-08-08, FERMILAB-PUB-09-855-T, BN-TH-2008-12 and SCIPP-08/08",
    doi = "10.1016/j.physrep.2010.05.002",
    journal = "Phys. Rept.",
    volume = "494",
    pages = "1--196",
    year = "2010"
}

@article{Nath:2006ut,
    author = "Nath, Pran and Fileviez Perez, Pavel",
    title = "{Proton stability in grand unified theories, in strings and in branes}",
    eprint = "hep-ph/0601023",
    archivePrefix = "arXiv",
    doi = "10.1016/j.physrep.2007.02.010",
    journal = "Phys. Rept.",
    volume = "441",
    pages = "191--317",
    year = "2007"
}

@article{Beneito:2023xbk,
    author = "Beneito, I, Arnau Bas and Gargalionis, John and Herrero-Garcia, Juan and Santamaria, Arcadi and Schmidt, Michael A.",
    title = "{An EFT approach to baryon number violation: lower limits on the new physics scale and correlations between nucleon decay modes}",
    eprint = "2312.13361",
    archivePrefix = "arXiv",
    primaryClass = "hep-ph",
    reportNumber = "IFIC/23-52, CPPC-2023-12",
    doi = "10.1007/JHEP07(2024)004",
    journal = "JHEP",
    volume = "07",
    pages = "004",
    year = "2024"
}

@article{Gargalionis:2024nij,
    author = "Gargalionis, John and Herrero-Garc\'\i{}a, Juan and Schmidt, Michael A.",
    title = "{Model-independent estimates for loop-induced baryon-number-violating nucleon decays}",
    eprint = "2401.04768",
    archivePrefix = "arXiv",
    primaryClass = "hep-ph",
    reportNumber = "CPPC-2024-02",
    doi = "10.1007/JHEP06(2024)182",
    journal = "JHEP",
    volume = "06",
    pages = "182",
    year = "2024"
}

@article{Buchmuller:1985jz,
    author = "Buchmuller, W. and Wyler, D.",
    title = "{Effective Lagrangian Analysis of New Interactions and Flavor Conservation}",
    reportNumber = "CERN-TH-4254/85",
    doi = "10.1016/0550-3213(86)90262-2",
    journal = "Nucl. Phys. B",
    volume = "268",
    pages = "621--653",
    year = "1986"
}

@article{Grzadkowski:2010es,
    author = "Grzadkowski, B. and Iskrzynski, M. and Misiak, M. and Rosiek, J.",
    title = "{Dimension-Six Terms in the Standard Model Lagrangian}",
    eprint = "1008.4884",
    archivePrefix = "arXiv",
    primaryClass = "hep-ph",
    reportNumber = "IFT-9-2010, TTP10-35",
    doi = "10.1007/JHEP10(2010)085",
    journal = "JHEP",
    volume = "10",
    pages = "085",
    year = "2010"
}

@article{Alonso:2014zka,
    author = "Alonso, Rodrigo and Chang, Hsi-Ming and Jenkins, Elizabeth E. and Manohar, Aneesh V. and Shotwell, Brian",
    title = "{Renormalization group evolution of dimension-six baryon number violating operators}",
    eprint = "1405.0486",
    archivePrefix = "arXiv",
    primaryClass = "hep-ph",
    doi = "10.1016/j.physletb.2014.05.065",
    journal = "Phys. Lett. B",
    volume = "734",
    pages = "302--307",
    year = "2014"
}

@article{Banik:2025wpi,
    author = "Banik, Sumit and Crivellin, Andreas and Naterop, Luca and Stoffer, Peter",
    title = "{Two-loop anomalous dimensions for baryon-number-violating operators in SMEFT}",
    eprint = "2510.08682",
    archivePrefix = "arXiv",
    primaryClass = "hep-ph",
    reportNumber = "ZU-TH 61/25",
    month = "10",
    year = "2025"
}

@article{Abbott:1980zj,
    author = "Abbott, L. F. and Wise, Mark B.",
    title = "{The Effective Hamiltonian for Nucleon Decay}",
    reportNumber = "SLAC-PUB-2487",
    doi = "10.1103/PhysRevD.22.2208",
    journal = "Phys. Rev. D",
    volume = "22",
    pages = "2208",
    year = "1980"
}

@article{Jenkins:2017dyc,
    author = "Jenkins, Elizabeth E. and Manohar, Aneesh V. and Stoffer, Peter",
    title = "{Low-Energy Effective Field Theory below the Electroweak Scale: Anomalous Dimensions}",
    eprint = "1711.05270",
    archivePrefix = "arXiv",
    primaryClass = "hep-ph",
    doi = "10.1007/JHEP01(2018)084",
    journal = "JHEP",
    volume = "01",
    pages = "084",
    year = "2018",
    note = "[Erratum: JHEP 12, 042 (2023)]"
}

@article{Aebischer:2025hsx,
    author = "Aebischer, Jason and Morell, Pol and Pesut, Marko and Virto, Javier",
    title = "{Two-Loop Anomalous Dimensions in the LEFT: Dimension-Six Four-Fermion Operators in NDR}",
    eprint = "2501.08384",
    archivePrefix = "arXiv",
    primaryClass = "hep-ph",
    month = "1",
    year = "2025"
}

@article{Naterop:2025lzc,
    author = "Naterop, Luca and Stoffer, Peter",
    title = "{Renormalization-group equations of the LEFT at two loops: dimension-six baryon-number-violating operators}",
    eprint = "2505.03871",
    archivePrefix = "arXiv",
    primaryClass = "hep-ph",
    reportNumber = "PSI-PR-25-09, ZU-TH 31/25, INT-PUB-25-013",
    doi = "10.1007/JHEP07(2025)237",
    journal = "JHEP",
    volume = "07",
    pages = "237",
    year = "2025"
}

@article{Jenkins:2017jig,
    author = "Jenkins, Elizabeth E. and Manohar, Aneesh V. and Stoffer, Peter",
    title = "{Low-Energy Effective Field Theory below the Electroweak Scale: Operators and Matching}",
    eprint = "1709.04486",
    archivePrefix = "arXiv",
    primaryClass = "hep-ph",
    doi = "10.1007/JHEP03(2018)016",
    journal = "JHEP",
    volume = "03",
    pages = "016",
    year = "2018",
    note = "[Erratum: JHEP 12, 043 (2023)]"
}

@article{Dekens:2019ept,
    author = "Dekens, Wouter and Stoffer, Peter",
    title = "{Low-energy effective field theory below the electroweak scale: matching at one loop}",
    eprint = "1908.05295",
    archivePrefix = "arXiv",
    primaryClass = "hep-ph",
    doi = "10.1007/JHEP10(2019)197",
    journal = "JHEP",
    volume = "10",
    pages = "197",
    year = "2019",
    note = "[Erratum: JHEP 11, 148 (2022)]"
}

@article{Aoki:2006ib,
    author = "Aoki, Y. and Dawson, C. and Noaki, J. and Soni, A.",
    title = "{Proton decay matrix elements with domain-wall fermions}",
    eprint = "hep-lat/0607002",
    archivePrefix = "arXiv",
    reportNumber = "BNL-HET-6-5, RBRC-606",
    doi = "10.1103/PhysRevD.75.014507",
    journal = "Phys. Rev. D",
    volume = "75",
    pages = "014507",
    year = "2007"
}

@article{Pivovarov:1991nk,
    author = "Pivovarov, A. A. and Surguladze, L. R.",
    title = "{Anomalous dimensions of octet baryonic currents in two loop approximation}",
    doi = "10.1016/0550-3213(91)90436-2",
    journal = "Nucl. Phys. B",
    volume = "360",
    pages = "97--108",
    year = "1991"
}

@article{Gracey:2012gx,
    author = "Gracey, J. A.",
    title = "{Three loop renormalization of 3-quark operators in QCD}",
    eprint = "1208.5619",
    archivePrefix = "arXiv",
    primaryClass = "hep-ph",
    reportNumber = "LTH-957",
    doi = "10.1007/JHEP09(2012)052",
    journal = "JHEP",
    volume = "09",
    pages = "052",
    year = "2012"
}

@article{JLQCD:1999dld,
    author = "Aoki, S. and others",
    collaboration = "JLQCD",
    title = "{Nucleon decay matrix elements from lattice QCD}",
    eprint = "hep-lat/9911026",
    archivePrefix = "arXiv",
    reportNumber = "HUPD-9919",
    doi = "10.1103/PhysRevD.62.014506",
    journal = "Phys. Rev. D",
    volume = "62",
    pages = "014506",
    year = "2000"
}

@article{Yoo:2021gql,
    author = "Yoo, Jun-Sik and Aoki, Yasumichi and Boyle, Peter and Izubuchi, Taku and Soni, Amarjit and Syritsyn, Sergey",
    title = "{Proton decay matrix elements on the lattice at physical pion mass}",
    eprint = "2111.01608",
    archivePrefix = "arXiv",
    primaryClass = "hep-lat",
    reportNumber = "RBRC-1333, KEK-CP-0385",
    doi = "10.1103/PhysRevD.105.074501",
    journal = "Phys. Rev. D",
    volume = "105",
    number = "7",
    pages = "074501",
    year = "2022"
}

@article{Aoki:2008ku,
    author = "Aoki, Y. and Boyle, P. and Cooney, P. and Del Debbio, L. and Kenway, R. and Maynard, C. M. and Soni, A. and Tweedie, R.",
    collaboration = "RBC-UKQCD",
    title = "{Proton lifetime bounds from chirally symmetric lattice QCD}",
    eprint = "0806.1031",
    archivePrefix = "arXiv",
    primaryClass = "hep-lat",
    reportNumber = "EDINBURGH-2008-19, RBRC-728, BNL-HET-08-13",
    doi = "10.1103/PhysRevD.78.054505",
    journal = "Phys. Rev. D",
    volume = "78",
    pages = "054505",
    year = "2008"
}

@article{Bali:2022qja,
    author = {Bali, Gunnar S. and Collins, Sara and S\"oldner, Wolfgang and Weish\"aupl, Simon},
    collaboration = "RQCD",
    title = "{Leading order mesonic and baryonic SU(3) low energy constants from Nf=3 lattice QCD}",
    eprint = "2201.05591",
    archivePrefix = "arXiv",
    primaryClass = "hep-lat",
    doi = "10.1103/PhysRevD.105.054516",
    journal = "Phys. Rev. D",
    volume = "105",
    number = "5",
    pages = "054516",
    year = "2022"
}

@article{Hyper-Kamiokande:2018ofw,
    author = "Abe, K. and others",
    collaboration = "Hyper-Kamiokande",
    title = "{Hyper-Kamiokande Design Report}",
    eprint = "1805.04163",
    archivePrefix = "arXiv",
    primaryClass = "physics.ins-det",
    month = "5",
    year = "2018"
}

@article{Super-Kamiokande:2005lev,
    author = "Kobayashi, K. and others",
    collaboration = "Super-Kamiokande",
    title = "{Search for nucleon decay via modes favored by supersymmetric grand unification models in Super-Kamiokande-I}",
    eprint = "hep-ex/0502026",
    archivePrefix = "arXiv",
    doi = "10.1103/PhysRevD.72.052007",
    journal = "Phys. Rev. D",
    volume = "72",
    pages = "052007",
    year = "2005"
}

@article{Super-Kamiokande:2012zik,
    author = "Regis, C. and others",
    collaboration = "Super-Kamiokande",
    title = "{Search for Proton Decay via $p -> \mu^+ K^0$ in Super-Kamiokande I, II, and III}",
    eprint = "1205.6538",
    archivePrefix = "arXiv",
    primaryClass = "hep-ex",
    doi = "10.1103/PhysRevD.86.012006",
    journal = "Phys. Rev. D",
    volume = "86",
    pages = "012006",
    year = "2012"
}

@article{Super-Kamiokande:2013rwg,
    author = "Abe, K. and others",
    collaboration = "Super-Kamiokande",
    title = "{Search for Nucleon Decay via $n \to \bar{\nu} \pi^{0}$ and $p \to \bar{\nu} \pi^{+}$ in Super-Kamiokande}",
    eprint = "1305.4391",
    archivePrefix = "arXiv",
    primaryClass = "hep-ex",
    doi = "10.1103/PhysRevLett.113.121802",
    journal = "Phys. Rev. Lett.",
    volume = "113",
    number = "12",
    pages = "121802",
    year = "2014"
}

@article{Super-Kamiokande:2014otb,
    author = "Abe, K. and others",
    collaboration = "Super-Kamiokande",
    title = "{Search for proton decay via $p\to\nu K^+$ using 260  kiloton{\textperiodcentered}year data of Super-Kamiokande}",
    eprint = "1408.1195",
    archivePrefix = "arXiv",
    primaryClass = "hep-ex",
    doi = "10.1103/PhysRevD.90.072005",
    journal = "Phys. Rev. D",
    volume = "90",
    number = "7",
    pages = "072005",
    year = "2014"
}

@article{Super-Kamiokande:2017gev,
    author = "Abe, K. and others",
    collaboration = "Super-Kamiokande",
    title = "{Search for nucleon decay into charged antilepton plus meson in 0.316 megaton$\cdot$years exposure of the Super-Kamiokande water Cherenkov detector}",
    eprint = "1705.07221",
    archivePrefix = "arXiv",
    primaryClass = "hep-ex",
    doi = "10.1103/PhysRevD.96.012003",
    journal = "Phys. Rev. D",
    volume = "96",
    number = "1",
    pages = "012003",
    year = "2017"
}

@article{Super-Kamiokande:2020wjk,
    author = "Takenaka, A. and others",
    collaboration = "Super-Kamiokande",
    title = "{Search for proton decay via $p\to e^+\pi^0$ and $p\to \mu^+\pi^0$ with an enlarged fiducial volume in Super-Kamiokande I-IV}",
    eprint = "2010.16098",
    archivePrefix = "arXiv",
    primaryClass = "hep-ex",
    doi = "10.1103/PhysRevD.102.112011",
    journal = "Phys. Rev. D",
    volume = "102",
    number = "11",
    pages = "112011",
    year = "2020"
}

@article{Super-Kamiokande:2022egr,
    author = "Matsumoto, R. and others",
    collaboration = "Super-Kamiokande",
    title = "{Search for proton decay via $p\rightarrow \mu^+K^0$ in 0.37 megaton-years exposure of Super-Kamiokande}",
    eprint = "2208.13188",
    archivePrefix = "arXiv",
    primaryClass = "hep-ex",
    doi = "10.1103/PhysRevD.106.072003",
    journal = "Phys. Rev. D",
    volume = "106",
    number = "7",
    pages = "072003",
    year = "2022"
}

@article{Super-Kamiokande:2024qbv,
    author = "Taniuchi, N. and others",
    collaboration = "Super-Kamiokande",
    title = "{Search for proton decay via p{\textrightarrow}e+{\ensuremath{\eta}} and p{\textrightarrow}{\ensuremath{\mu}}+{\ensuremath{\eta}} with a 0.37~Mton-year exposure of Super-Kamiokande}",
    eprint = "2409.19633",
    archivePrefix = "arXiv",
    primaryClass = "hep-ex",
    doi = "10.1103/PhysRevD.110.112011",
    journal = "Phys. Rev. D",
    volume = "110",
    number = "11",
    pages = "112011",
    year = "2024"
}

@article{Super-Kamiokande:2025ibz,
    author = "Yamauchi, K. and others",
    collaboration = "Super-Kamiokande",
    title = "{Search for neutron decay into an antineutrino and a neutral kaon in 0.401 megaton-years exposure of Super-Kamiokande}",
    eprint = "2506.14406",
    archivePrefix = "arXiv",
    primaryClass = "hep-ex",
    month = "6",
    year = "2025"
}

@article{Minkowski:1977sc,
    author = "Minkowski, Peter",
    title = "{$\mu \to e\gamma$ at a Rate of One Out of $10^{9}$ Muon Decays?}",
    reportNumber = "Print-77-0182 (BERN)",
    doi = "10.1016/0370-2693(77)90435-X",
    journal = "Phys. Lett. B",
    volume = "67",
    pages = "421--428",
    year = "1977"
}

@article{Yanagida:1979as,
    author = "Yanagida, Tsutomu",
    editor = "Sawada, Osamu and Sugamoto, Akio",
    title = "{Horizontal gauge symmetry and masses of neutrinos}",
    reportNumber = "KEK-79-18-95",
    journal = "Conf. Proc. C",
    volume = "7902131",
    pages = "95--99",
    year = "1979"
}

@article{Gell-Mann:1979vob,
    author = "Gell-Mann, Murray and Ramond, Pierre and Slansky, Richard",
    title = "{Complex Spinors and Unified Theories}",
    eprint = "1306.4669",
    archivePrefix = "arXiv",
    primaryClass = "hep-th",
    reportNumber = "PRINT-80-0576",
    journal = "Conf. Proc. C",
    volume = "790927",
    pages = "315--321",
    year = "1979"
}

@article{Mohapatra:1979ia,
    author = "Mohapatra, Rabindra N. and Senjanovic, Goran",
    title = "{Neutrino Mass and Spontaneous Parity Nonconservation}",
    reportNumber = "MDDP-TR-80-060, MDDP-PP-80-105, CCNY-HEP-79-10",
    doi = "10.1103/PhysRevLett.44.912",
    journal = "Phys. Rev. Lett.",
    volume = "44",
    pages = "912",
    year = "1980"
}

@article{Weinberg:1979sa,
    author = "Weinberg, Steven",
    title = "{Baryon and Lepton Nonconserving Processes}",
    reportNumber = "HUTP-79-A050",
    doi = "10.1103/PhysRevLett.43.1566",
    journal = "Phys. Rev. Lett.",
    volume = "43",
    pages = "1566--1570",
    year = "1979"
}

@article{Kaplan:1991dc,
    author = "Kaplan, David B.",
    title = "{Flavor at SSC energies: A New mechanism for dynamically generated fermion masses}",
    reportNumber = "UCSD-PTH-91-04",
    doi = "10.1016/S0550-3213(05)80021-5",
    journal = "Nucl. Phys. B",
    volume = "365",
    pages = "259--278",
    year = "1991"
}

@article{Kaplan:1983fs,
    author = "Kaplan, David B. and Georgi, Howard",
    title = "{SU(2) x U(1) Breaking by Vacuum Misalignment}",
    reportNumber = "HUTP-83/A069",
    doi = "10.1016/0370-2693(84)91177-8",
    journal = "Phys. Lett. B",
    volume = "136",
    pages = "183--186",
    year = "1984"
}

@article{Kaplan:1983sm,
    author = "Kaplan, David B. and Georgi, Howard and Dimopoulos, Savas",
    title = "{Composite Higgs Scalars}",
    reportNumber = "HUTP-83/A079",
    doi = "10.1016/0370-2693(84)91178-X",
    journal = "Phys. Lett. B",
    volume = "136",
    pages = "187--190",
    year = "1984"
}

@article{Georgi:1984ef,
    author = "Georgi, Howard and Kaplan, David B. and Galison, Peter",
    title = "{Calculation of the Composite Higgs Mass}",
    reportNumber = "HUTP-84/A018",
    doi = "10.1016/0370-2693(84)90823-2",
    journal = "Phys. Lett. B",
    volume = "143",
    pages = "152--154",
    year = "1984"
}

@article{Georgi:1984af,
    author = "Georgi, Howard and Kaplan, David B.",
    title = "{Composite Higgs and Custodial SU(2)}",
    reportNumber = "HUTP-84/A034",
    doi = "10.1016/0370-2693(84)90341-1",
    journal = "Phys. Lett. B",
    volume = "145",
    pages = "216--220",
    year = "1984"
}

@article{Dugan:1984hq,
    author = "Dugan, Michael J. and Georgi, Howard and Kaplan, David B.",
    title = "{Anatomy of a Composite Higgs Model}",
    reportNumber = "HUTP-84/A073",
    doi = "10.1016/0550-3213(85)90221-4",
    journal = "Nucl. Phys. B",
    volume = "254",
    pages = "299--326",
    year = "1985"
}

@article{Contino:2003ve,
    author = "Contino, Roberto and Nomura, Yasunori and Pomarol, Alex",
    title = "{Higgs as a Holographic Pseudo Goldstone Boson}",
    eprint = "hep-ph/0306259",
    archivePrefix = "arXiv",
    reportNumber = "FT-UAM-03-11, FERMILAB-PUB-03-195-T, UAB-FT-549",
    doi = "10.1016/j.nuclphysb.2003.08.027",
    journal = "Nucl. Phys. B",
    volume = "671",
    pages = "148--174",
    year = "2003"
}

@article{Degrassi:2012ry,
    author = "Degrassi, Giuseppe and Di Vita, Stefano and Elias-Miro, Joan and Espinosa, Jose R. and Giudice, Gian F. and Isidori, Gino and Strumia, Alessandro",
    title = "{Higgs mass and vacuum stability in the Standard Model at NNLO}",
    eprint = "1205.6497",
    archivePrefix = "arXiv",
    primaryClass = "hep-ph",
    reportNumber = "CERN-PH-TH-2012-134, RM3-TH-12-9",
    doi = "10.1007/JHEP08(2012)098",
    journal = "JHEP",
    volume = "08",
    pages = "098",
    year = "2012"
}

@article{Hiller:2024zjp,
    author = {Hiller, Gudrun and H{\"o}hne, Tim and Litim, Daniel F. and Steudtner, Tom},
    title = "{Vacuum stability in the Standard Model and beyond}",
    eprint = "2401.08811",
    archivePrefix = "arXiv",
    primaryClass = "hep-ph",
    doi = "10.1103/PhysRevD.110.115017",
    journal = "Phys. Rev. D",
    volume = "110",
    number = "11",
    pages = "115017",
    year = "2024"
}

@article{Froggatt:1978nt,
    author = "Froggatt, C. D. and Nielsen, Holger Bech",
    title = "{Hierarchy of Quark Masses, Cabibbo Angles and CP Violation}",
    reportNumber = "CERN-TH-2519",
    doi = "10.1016/0550-3213(79)90316-X",
    journal = "Nucl. Phys. B",
    volume = "147",
    pages = "277--298",
    year = "1979"
}

@article{Huber:2003tu,
    author = "Huber, Stephan J.",
    title = "{Flavor violation and warped geometry}",
    eprint = "hep-ph/0303183",
    archivePrefix = "arXiv",
    reportNumber = "DESY-03-037",
    doi = "10.1016/S0550-3213(03)00502-9",
    journal = "Nucl. Phys. B",
    volume = "666",
    pages = "269--288",
    year = "2003"
}

@article{Csaki:2008zd,
    author = "Csaki, Csaba and Falkowski, Adam and Weiler, Andreas",
    title = "{The Flavor of the Composite Pseudo-Goldstone Higgs}",
    eprint = "0804.1954",
    archivePrefix = "arXiv",
    primaryClass = "hep-ph",
    doi = "10.1088/1126-6708/2008/09/008",
    journal = "JHEP",
    volume = "09",
    pages = "008",
    year = "2008"
}

@article{Ibe:2024cbt,
    author = "Ibe, Masahiro and Shirai, Satoshi and Watanabe, Keiichi",
    title = "{Nucleon decay as a probe of flavor symmetry: the case of fake unification}",
    eprint = "2411.05398",
    archivePrefix = "arXiv",
    primaryClass = "hep-ph",
    reportNumber = "IPMU24-0039",
    doi = "10.1007/JHEP03(2025)044",
    journal = "JHEP",
    volume = "03",
    pages = "044",
    year = "2025"
}

@article{Ibe:2024cvi,
    author = "Ibe, Masahiro and Shirai, Satoshi and Watanabe, Keiichi",
    title = "{Comprehensive Bayesian exploration of Froggatt-Nielsen mechanism}",
    eprint = "2412.19484",
    archivePrefix = "arXiv",
    primaryClass = "hep-ph",
    reportNumber = "IPMU24-0047",
    doi = "10.1007/JHEP03(2025)150",
    journal = "JHEP",
    volume = "03",
    pages = "150",
    year = "2025"
}

@article{Chitose:2025bvl,
    author = "Chitose, Akifumi and Ibe, Masahiro and Shirai, Satoshi",
    title = "{Flavor Symmetry and Proton Decay in PeV-Scale Supersymmetry}",
    eprint = "2510.13617",
    archivePrefix = "arXiv",
    primaryClass = "hep-ph",
    reportNumber = "IPMU25-0049",
    month = "10",
    year = "2025"
}

@article{Beneito:2025fzf,
    author = "Beneito, I, Arnau Bas and Palavri{\'c}, Ajdin and Sainaghi, Andrea",
    title = "{Implications of Flavor Symmetries for Baryon Number Violation}",
    eprint = "2512.14813",
    archivePrefix = "arXiv",
    primaryClass = "hep-ph",
    month = "12",
    year = "2025"
}

@article{Co:2019wyp,
    author = "Co, Raymond T. and Harigaya, Keisuke",
    title = "{Axiogenesis}",
    eprint = "1910.02080",
    archivePrefix = "arXiv",
    primaryClass = "hep-ph",
    reportNumber = "LCTP-19-27",
    doi = "10.1103/PhysRevLett.124.111602",
    journal = "Phys. Rev. Lett.",
    volume = "124",
    number = "11",
    pages = "111602",
    year = "2020"
}

@article{Co:2019jts,
    author = "Co, Raymond T. and Hall, Lawrence J. and Harigaya, Keisuke",
    title = "{Axion Kinetic Misalignment Mechanism}",
    eprint = "1910.14152",
    archivePrefix = "arXiv",
    primaryClass = "hep-ph",
    reportNumber = "LCTP-19-28",
    doi = "10.1103/PhysRevLett.124.251802",
    journal = "Phys. Rev. Lett.",
    volume = "124",
    number = "25",
    pages = "251802",
    year = "2020"
}

@article{Co:2020dya,
    author = "Co, Raymond T. and Hall, Lawrence J. and Harigaya, Keisuke and Olive, Keith A. and Verner, Sarunas",
    title = "{Axion Kinetic Misalignment and Parametric Resonance from Inflation}",
    eprint = "2004.00629",
    archivePrefix = "arXiv",
    primaryClass = "hep-ph",
    reportNumber = "LCTP-20-06, UMN-TH-3912/20, FTPI-MINN-20/02",
    doi = "10.1088/1475-7516/2020/08/036",
    journal = "JCAP",
    volume = "08",
    pages = "036",
    year = "2020"
}

@article{Domcke:2020kcp,
    author = "Domcke, Valerie and Ema, Yohei and Mukaida, Kyohei and Yamada, Masaki",
    title = "{Spontaneous Baryogenesis from Axions with Generic Couplings}",
    eprint = "2006.03148",
    archivePrefix = "arXiv",
    primaryClass = "hep-ph",
    reportNumber = "DESY 20-100, DESY-20-100, CERN-TH-2020-088, TU-1102",
    doi = "10.1007/JHEP08(2020)096",
    journal = "JHEP",
    volume = "08",
    pages = "096",
    year = "2020"
}

@article{Chivukula:1987py,
    author = "Chivukula, R. Sekhar and Georgi, Howard",
    title = "{Composite Technicolor Standard Model}",
    reportNumber = "BUHEP-87-2, HUTP-87/A003",
    doi = "10.1016/0370-2693(87)90713-1",
    journal = "Phys. Lett. B",
    volume = "188",
    pages = "99--104",
    year = "1987"
}

@article{DAmbrosio:2002vsn,
    author = "D'Ambrosio, G. and Giudice, G. F. and Isidori, G. and Strumia, A.",
    title = "{Minimal flavor violation: An Effective field theory approach}",
    eprint = "hep-ph/0207036",
    archivePrefix = "arXiv",
    reportNumber = "CERN-TH-2002-147, IFUP-TH-2002-17",
    doi = "10.1016/S0550-3213(02)00836-2",
    journal = "Nucl. Phys. B",
    volume = "645",
    pages = "155--187",
    year = "2002"
}

@article{Barbieri:2011ci,
    author = "Barbieri, Riccardo and Isidori, Gino and Jones-Perez, Joel and Lodone, Paolo and Straub, David M.",
    title = "{$U(2)$ and Minimal Flavour Violation in Supersymmetry}",
    eprint = "1105.2296",
    archivePrefix = "arXiv",
    primaryClass = "hep-ph",
    doi = "10.1140/epjc/s10052-011-1725-z",
    journal = "Eur. Phys. J. C",
    volume = "71",
    pages = "1725",
    year = "2011"
}

@article{Barbieri:2012uh,
    author = "Barbieri, Riccardo and Buttazzo, Dario and Sala, Filippo and Straub, David M.",
    title = "{Flavour physics from an approximate $U(2)^3$ symmetry}",
    eprint = "1203.4218",
    archivePrefix = "arXiv",
    primaryClass = "hep-ph",
    doi = "10.1007/JHEP07(2012)181",
    journal = "JHEP",
    volume = "07",
    pages = "181",
    year = "2012"
}

@article{Blankenburg:2012nx,
    author = "Blankenburg, Gianluca and Isidori, Gino and Jones-Perez, Joel",
    title = "{Neutrino Masses and LFV from Minimal Breaking of $U(3)^5$ and $U(2)^5$ flavor Symmetries}",
    eprint = "1204.0688",
    archivePrefix = "arXiv",
    primaryClass = "hep-ph",
    reportNumber = "CERN-PH-TH-2012-082",
    doi = "10.1140/epjc/s10052-012-2126-7",
    journal = "Eur. Phys. J. C",
    volume = "72",
    pages = "2126",
    year = "2012"
}

@article{Nikolidakis:2007fc,
    author = "Nikolidakis, Emanuel and Smith, Christopher",
    title = "{Minimal Flavor Violation, Seesaw, and R-parity}",
    eprint = "0710.3129",
    archivePrefix = "arXiv",
    primaryClass = "hep-ph",
    doi = "10.1103/PhysRevD.77.015021",
    journal = "Phys. Rev. D",
    volume = "77",
    pages = "015021",
    year = "2008"
}

@article{Smith:2011rp,
    author = "Smith, Christopher",
    title = "{Proton stability from a fourth family}",
    eprint = "1105.1723",
    archivePrefix = "arXiv",
    primaryClass = "hep-ph",
    doi = "10.1103/PhysRevD.85.036005",
    journal = "Phys. Rev. D",
    volume = "85",
    pages = "036005",
    year = "2012"
}

@article{Helset:2019eyc,
    author = "Helset, Andreas and Kobach, Andrew",
    title = "{Baryon Number, Lepton Number, and Operator Dimension in the SMEFT with Flavor Symmetries}",
    eprint = "1909.05853",
    archivePrefix = "arXiv",
    primaryClass = "hep-ph",
    doi = "10.1016/j.physletb.2019.135132",
    journal = "Phys. Lett. B",
    volume = "800",
    pages = "135132",
    year = "2020"
}

\end{document}